\documentclass[12pt,a4paper]{article}

\usepackage{amsmath,amssymb}
\usepackage[bookmarks=true]{hyperref}
\usepackage[nosort]{cite}
\usepackage[bulletsep]{collect}
\def\gfxon{\usepackage[final]{graphicx}}

\gfxon

\makeatletter
\let\old@startsection=\@startsection
\renewcommand{\@startsection}[6]{\old@startsection{#1}{#2}{#3}{#4}{#5}{#6\mathversion{bold}}}
\makeatother

\makeatletter \@addtoreset{equation}{section} \makeatother

\makeatletter
\let\old@makecaption=\@makecaption
\def\@makecaption{\small\old@makecaption}
\makeatother

\newcommand{\ham}{\mathcal{H}}

\newcommand{\gen}[1]{\mathfrak{#1}}

\makeatletter
\newcommand{\ellSN}{\mathop{\operator@font sn}\nolimits}
\newcommand{\ellCN}{\mathop{\operator@font cn}\nolimits}
\newcommand{\ellDN}{\mathop{\operator@font dn}\nolimits}
\newcommand{\ellAM}{\mathop{\operator@font am}\nolimits}
\newcommand{\ellK}{\mathop{\smash{\operator@font K}\vphantom{a}}\nolimits}
\newcommand{\ellE}{\mathop{\smash{\operator@font E}\vphantom{a}}\nolimits}
\makeatother

\ifx\genfrac\sdflkaj

\else

\fi
\newcommand{\sfrac}[2]{{\textstyle\frac{#1}{#2}}}
\newcommand{\half}{\sfrac{1}{2}}

\newcommand{\vev}[1]{\langle#1\rangle}

\newcommand{\ssN}{\mathcal{N}}

\newcommand{\alg}[1]{\mathfrak{#1}}

\newcommand{\nln}{\nonumber\\}
\newcommand{\nl}[1][0pt]{\nonumber\\[#1]&\hspace{-4\arraycolsep}&\mathord{}}

\newcommand{\earel}[1]{\mathrel{}&\hspace{-2\arraycolsep}#1\hspace{-2\arraycolsep}&\mathrel{}}
\newcommand{\eq}{\earel{=}}
\newcommand{\beq}{\begin{equation}}
\newcommand{\eeq}{\end{equation}}

\def\[{\begin{equation}}
\def\]{\end{equation}}
\def\<{\begin{eqnarray}}
\def\>{\end{eqnarray}}

\makeatletter
\def\mr@ignsp#1 {\ifx\:#1\@empty\else #1\expandafter\mr@ignsp\fi}%
\newcommand{\multiref}[1]{\begingroup
\xdef\mr@no@sparg{\expandafter\mr@ignsp#1 \: }%
\def\mr@comma{}%
\@for\mr@refs:=\mr@no@sparg\do{\mr@comma\def\mr@comma{,}\ref{\mr@refs}}%
\endgroup}
\makeatother

\newcommand{\hypref}[2]{\ifx\href\asklfhas #2\else\href{#1}{#2}\fi}
\newcommand{\Secref}[1]{Section~\multiref{#1}}
\newcommand{\secref}[1]{Sec.~\multiref{#1}}

\newcommand{\tabref}[1]{Tab.~\multiref{#1}}

\newcommand{\figref}[1]{Fig.~\multiref{#1}}
\renewcommand{\eqref}[1]{(\multiref{#1})}

\ifx\href\asklfhas\newcommand{\href}[2]{#2}\fi

\begin{document}

\begin{flushright}\footnotesize
\vspace{0.5cm}
\end{flushright}
\vspace{0.3cm}

\renewcommand{\thefootnote}{\arabic{footnote}}

\setcounter{footnote}{0}

\begin{center}%

\vspace{-.7cm}%

{\Large\textbf{\mathversion{bold}
On the Integrability of Four Dimensional $\ssN=2$ Gauge Theories in the Omega Background} 
\par}


\vspace{8mm}

\thispagestyle{empty}

\textsc{Heng-Yu Chen$^{1}$, Po-Shen Hsin$^{1}$ and Peter Koroteev$^{2}$}

\vspace{4mm}%

\textit{$^{1}$Department of Physics and Center for Theoretical Sciences\\%
National Taiwan University, Taipei 10617, Taiwan}

\vspace{3mm}%

\textit{$^{2}$University of Minnesota, School of Physics and Astronomy\\%
116 Church Street S.E. Minneapolis, MN 55455, USA}

\vspace{3mm}%

\textit{$^{2}$Perimeter Institute for Theoretical Physics\\%
31 Caroline Street North, ON N2L2Y5, Canada}

\vspace{8mm}

\texttt{heng.yu.chen@phys.ntu.edu.tw, nazgoulz@gmail.com,\\%
 pkoroteev@perimeterinstitute.ca}

\par\vspace{.5cm}

\vfill

\textbf{Abstract}\vspace{5mm}

\begin{minipage}{12.7cm}
We continue to investigate the relationship between the infrared physics of $\mathcal{N}=2$ supersymmetric gauge theories in four dimensions and various integrable models such as Gaudin, Calogero-Moser and quantum spin chains. We prove interesting dualities among some of these integrable systems by performing different, albeit equivalent, quantizations of the Seiberg-Witten curve of the four dimensional theory. We also discuss conformal field theories related to $\mathcal{N}=2$ 4d gauge theories by the Alday-Gaiotto-Tachikawa (AGT) duality and the role of conformal blocks of those CFTs in the integrable systems. As a consequence, the equivalence of conformal blocks of rank two Toda and Novikov-Wess-Zumino-Witten (WZNW) theories on the torus with punctures is found.
\end{minipage}

\vspace*{\fill}
\end{center}

\newpage

\tableofcontents


\section{Introduction}\label{sec:Intro}
Four dimensional supersymmetric gauge theories with $\ssN=2$ supersymmetry have been known for some time to be deeply related to various integrable systems. After the appearance of Seiberg-Witten solution \cite{Seiberg:1994rs,Seiberg:1994aj} many interesting connections between the infrared physics of $\ssN=2$ gauge theories and integrability \cite{Gorsky:1995zq,Donagi:1995cf} have emerged. In particular, the Seiberg-Witten (SW) curve of a gauge theory is mapped onto the spectral curve of the corresponding classical integrable model, such as the Hitchin system \cite{0887284,0627.14024}. Using string theory and M-theory constructions, it was possible to construct SW curves for a wide class of 4d theories from the UV data of the theory \cite{Witten:1997sc}. However, not all interesting theories can be constructed this way. Recently the solution \cite{Nekrasov:2012xe} of a wider class of conformal theories has been found by Nekrasov and Pestun, which includes all 4d quiver theories built out of unitary groups whose shapes coincide with Dynkin diagrams of ADE groups and their affine cousins. 

Having obtained a classical integrable system from a gauge theory, it is tempting to quantize it and understand what the quantization means in terms of the field theory we started with. A universal prescription of the quantization was proposed by Nekrasov and Shatashvili (NS) \cite{Nekrasov:2009rc}: the gauge theory in question is subject to the $\Omega$-deformation, with two parameters $\epsilon_1$ and $\epsilon_2$, which break the Lorentz invariance of the 4d theory. For the purposes of \cite{Nekrasov:2009rc}, as well as for this note, we take $\epsilon_2=0$, which is referred to as the \textit{NS limit}. The remaining parameter $\epsilon_1$ is mirrored in the integrable model as the Planck constant. 

Alternatively the integrable many-body system underlying a gauge theory can be quantized canonically and its spectrum can be found. One should expect the result to coincide with NS quantization we have just mentioned, in other words, the two quantizations of the same classical system must be equivalent. They could be merely identical, and if they are not, there should be a duality between them involved, which maps spectra of both models one to the other. In the literature such dualities are usually called \textit{bispectral} \cite{springerlink:10.1007/BF00626526,MR2409414,Mironov:2012ba, Mironov:2012uh, Bulycheva:2012ct} \footnote{In the literature there has been a lot of confusion about that term, sometimes \textit{spectral} is also used. Sometimes, however, the duality looks very different \cite{Gadde:2013wq}.}. 

In this note we shall consider an example of such equivalence emerging from 4d linear $A_{L}$ conformal quiver theories. 
For our purposes we will not need the whole Coulomb branch, rather special loci thereof -- roots of the Higgs branch of the theory, which gets slightly modified in the presence of the $\Omega$-deformation. At Higgs branch roots the SW curve greatly simplifies and, as it was shown by Dorey, Hollowood, Lee and one of current authors (CDHL) \cite{Dorey:2011pa,Chen:2011sj}, that in the NS limit the Nekrasov prepotential \cite{Nekrasov:2002qd} in the dual frame can be regarded as the effective twisted superpotential of a certain two dimensional $(2,2)$ gauged linear sigma model (GLSM). Vacuum equations of this sigma model are equivalent to Bethe ansatz equations of the quantum anisotropic XXX spin chain, whose symmetry group has rank $L$. Whereas the SW curve is the spectral curve of a Hitchin system on $S^2$ with $L+3$ punctures (also known as the Gaudin model \cite{garnier1976rend}). Its quantization leads to the eigenvalue problem of the Gaudin Hamiltonian. 
The latter can be canonically quantized and its spectrum can be reduced to the eigenvalue problem of the Gaudin Hamiltonian.
Diagonalization of both XXX \cite{Bethe:1931hc} and Gaudin \cite{refId0} Hamiltonians can be effectively done using the Bethe ansatz technique. Because of the aforementioned equivalence, the two models should be bispectrally dual to each other and the parameter spaces of solutions should be in a correspondence. 

In \cite{Mironov:2012ba} a proof for the $\mathfrak{gl}(N)/\mathfrak{gl}(M)$ XXX/Gaudin pair was provided both on the classical and on the quantum levels using the spectral curves of the models and Baxter equations for them. However, for the quantum duality the authors of \cite{Mironov:2012ba} did not specify the wave functions of the problem nor the boundary conditions for Baxter operators. Therefore it still remained to be unknown how the parameter spaces of solutions of the quantum problems are mapped onto each other. A step in this direction was made in \cite{MR2409414}: the bispectral duality was proven for rank one systems and conjectured for higher ranks using Bethe equations for both models. One of the goals of the current paper is to prove the most generic duality of this type using the string theory \cite{Gorsky:1997jq,Gorsky:1997mw}. 

Very recently in \cite{Gaiotto:2013bwa} the XXX/Gaudin duality was proven in its most generic form, albeit for \textit{compact} representations. It was shown that the bispectral duality follows from the mirror symmetry of certain $\ssN=2^*$ three dimensional quiver gauge theories\footnote{That is, we break supersymmetry partially by giving mass to the adjoint scalar in three dimensional ${\mathcal N}=4$ vector multiplet}. After compactifying the 3d theory on a circle, its vacua parameter space (masses, Fayet-Iliopoulos terms) could be interpreted as a space of solutions of the XXZ spin chain, where the anisotropy parameter was inversely proportional to the compactification radius. Thus each XXZ chain had its mirror (bispectral) dual XXZ$^\vee$. Remarkably, the XXX/Gaudin bispectral duality followed from the XXZ/XXZ$^\vee$ relation by taking the zero radius limit. Compactness of the symmetry group of the spin chain was crucial in the story, in fact it was tied up with the action of the diagonal $U(1)$ R-symmetry generator and relative charges of the matter fields under this R-symmetry. In addition to that the type-IIB brane construction, which was used in \cite{Gaiotto:2013bwa} to illustrate the 3d mirror symmetry via the S-duality, constrained the number of solutions of the XXZ Bethe equations with the help of the s-rule \cite{Hanany:1996ie}, which constrained the number of D2 branes in the construction, or, equivalently, the number of Bethe roots in the spin chain.
 
The situation with 4d quiver theories is, however, different from \cite{Gaiotto:2013bwa}. One should not expect any finiteness constraints on the Hilbert space of states emerging from the CDHL quantization due to several reasons. 

First, the data of this GLSM in \cite{Dorey:2011pa,Chen:2011sj} were partly provided by the UV description of the 4d quiver theory, and in part by the quantization at the roots of the Higgs branch of the theory (we shall call it the \textit{CDHL quantization}). The GLSM appearing from the CDHL quantization is generically non-abelian and the rank of the gauge group has a direct dependence on the quiver data and on the quantization conditions. These facts lead to the conclusion \cite{Dorey:2011pa,Chen:2011sj} that the ground state equations for the GLSM in question coincide with the XXX Bethe equations with symmetry $SL(K)$ for some $K$. 

Second, it was argued in \cite{Dorey:2011pa,Bulycheva:2012ct} that the worldvolume theory in the NS $\Omega$-background effectively becomes two-dimensional and can be viewed as the worldsheet of non-abelian vortex strings (see \cite{Shifman:2007ce} and references therein). The same kind of strings appeared as D2 branes in the CDHL construction, where the winding numbers of the strings were related to the number of the D2 branes and they gave the quantization for $a-m$ cycles of the SW curve of the 4d theory. Again, there is no \textit{a priori} constraint on the winding numbers, that the symmetry of the quantum many-body system is expected to be noncompact. We shall discuss these quantum systems in \Secref{sec:VerifBispec}.

The Alday-Gaiotto-Tachikawa (AGT) duality \cite{Alday:2009aq} between partition functions of 4d gauge theories and conformal blocks of 2d CFTs implies a certain relation between the equivariant parameters $\epsilon_1$ and $\epsilon_2$ of the 4d $\Omega$-background and the CFT central charge. If on top of this relation one imposes the NS limit, it drives the CFT into the semiclassical regime, since the central charge becomes large. Finally, if in addition to that one sits at the Higgs branch roots of the 4d theory, then conformal blocks of the dual CFT dramatically simplify since external momenta become degenerate. For instance, for $SU(2)$ SQCD the corresponding Liouville conformal block in the NS limit becomes an eigenfunction of the Gaudin Hamiltonian \cite{Bulycheva:2012ct}. We will comment about some generalizations of this fact in \Secref{sec:SpherePunct}.

Along with genus zero spectral curves for the Gaudin model we also discuss genus one curves and the corresponding Calogero-Moser \cite{MR0280103,MR0375869,Sutherland:1971ks} type systems in \Secref{sec:Torus}. This model appears in NS quantization of $\ssN=2^*$ theory in four dimensions. Its dual conformal theory via the AGT duality is the Toda theory on a single punctured torus. Another interesting CFT which turns out to be relevant in this discussion is the gauged Novikov-Wess-Zumino-Witten (WZNW) model which enjoys an affine algebra symmetry. We study in the details the rank two WZNW theory and show that its conformal blocks on a torus with one puncture satisfy the equation of motion for the elliptic Calogero system. As we shall show later, the same equation is solved by $\mathcal{W}_3$ conformal blocks in the NS limit. In order for the matching to occur the affine algebra level has to tuned to its critical value. Therefore we are able to generalize some results in the literature \cite{Ribault:2005wp,Ribault:2008si,Hikida:2007tq} for rank two CFTs on the torus.

The paper is organized as follows. In \Secref{sec:VerifBispec} we prove the bispectral duality using several results about the Seiberg-Witten geometry of quiver gauge theories. The conjectures from the mathematical literature are proven and generalized to more complicated representations. 
Then in \Secref{sec:Torus} we discuss rank two Toda and WZNW CFTs on the torus and their connection to integrable systems. Finally, \Secref{sec:Future} outlines some possible future research directions. 

\section{The Bispectral Duality from Supersymmetric Gauge Theories}\label{sec:VerifBispec}
Here we would like to generalize the result of \cite{Bulycheva:2012ct} to $A_L$ quiver gauge theories with arbitrary color and flavor labels $(N_1,M_1)\dots (N_L,M_L)$ which obey asymptotic conformal invariance on every gauge node: 
\begin{equation}
N_{i+1}+N_{i-1}+M_i-2N_i=0\,,
\label{eq:betafuncitonzero}
\end{equation}
which is the condition for all beta functions to vanish. In principle our analysis can be extended to other ADE-type quivers and their affine generalizations as in \cite{Nekrasov:2012xe}, however we shall postpone these to the future work, so let us focus on $A_L$ quivers. Evidently there are two major steps toward the generic quiver theory. First we need to look at linear $U(2)$ quiver for $L\geq 2$, then consider more generic gauge and flavor labels on each node provided that \eqref{eq:betafuncitonzero} holds. In this section we start with the first generalization.

\begin{figure}
\begin{center}
\includegraphics[width=14cm, height=6.5cm]{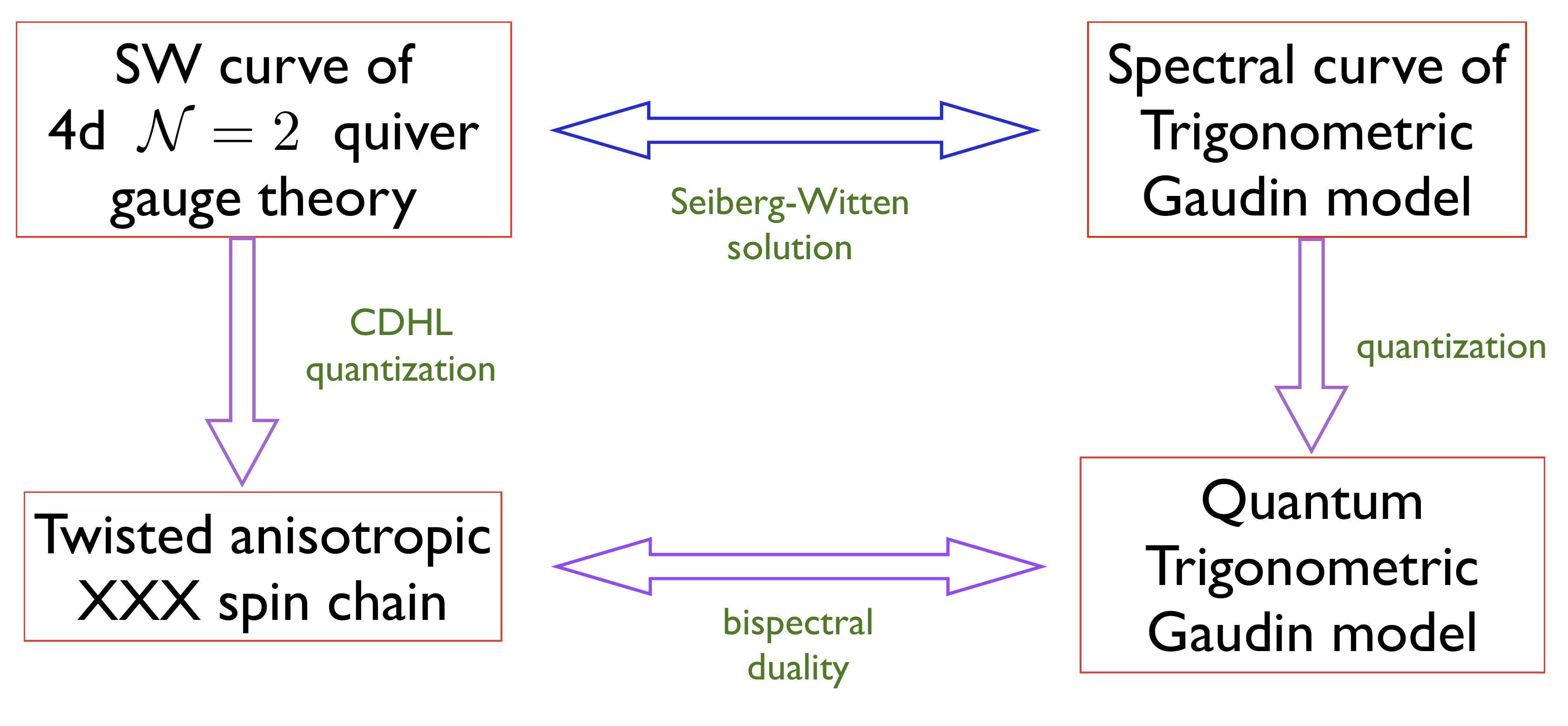}
\caption{Chain of dualities involving 4d and 2d quivers}
\label{fig:flowchartsphere}
\end{center}
\end{figure}

The main result of this section is to establish the so-called bispectral duality between two distinct quantum integrable systems, namely XXX spin chain and quantum Gaudin system, by investigating the exact vacua of supersymmetric gauge theories.
Our strategy is summarized in \figref{fig:flowchartsphere}, 
we will first invoke a couple of non-trivial correspondences established in \cite{Witten:1997sc,Nekrasov:2012xe,Gaiotto:2009hg} and \cite{Chen:2011sj} relating supersymmetric field theories to the two different quantum integrable systems considered then join the dots accordingly. 

\subsection{The $\mathfrak{sl}(N)/\mathfrak{sl}(2)$ Duality}
Let us begin by specifying in some details the supersymmetric gauge theories we are considering, 
our starting point is four dimensional ${\mathcal{N}}=2$ linear quiver theory with gauge group $\prod^{L}_{I=1} U(2)_I$, each gauge node has Coulomb branch coordinates $a^{(I)}_a$, $I=1, 2\dots, L, ~ a=1,2$ and complexified gauge coupling $q_I = e^{2\pi i \tau_I }$. The matter content of this theory consists of bi-fundamental hypermultiplets of masses $\mu_{I}$ connecting successive $I$th and $(I+1)$th gauge groups, plus two anti-fundamental and two fundamental multiplets of masses $m_{1,2}^{(1)}$ and $m_{1,2}^{(L)}$ respectively each of them charged under the $SU(2)$ flavor symmetry.
Here we shall subject this theory to a special limit of equivariant localization introduced by Nekrasov and Shatashvili \cite{Nekrasov:2009rc} (hence referred as ``NS limit'') when one of the two equivariant parameters of the 4d gauge theory, say $\epsilon_2$, vanishes. In such a limit, the key quantity for describing the low energy quantum dynamics of the system is given by:
\[
{\mathcal{F}}(a^{(I)}_a; \epsilon)=\lim_{(\epsilon_1,\epsilon_2) \to (\epsilon, 0)} \epsilon_1\epsilon_2\log {\mathcal Z}(a^{(I)}_a; \epsilon_1, \epsilon_2)
\label{eq:NSpart}
\] 
where ${\mathcal Z}(a^{(I)}_a; \epsilon_1, \epsilon_2)$ is the Nekrasov partition function, including both the perturbative and non-perturbative contributions.

The linear quiver gauge theory in the NS limit has been elaborated in \cite{Chen:2011sj}. It was shown that at its quantized root of the Higgs branch, which is specified by the following condition:
\[
a^{(I)}_a = m_a^{(L)} - n^{(I)}_a \epsilon - \sum\limits_{J=I}^L \mu_J\,, \quad a=1,2\,,
\label{eq:BarHiggs2}
\]
where $n^{(I)}_a$ are positive integers, there exist codimension two BPS topological solitons known as ``vortices'' whose winding number is precisely given by the flux quantization number $n^{(I)}_a$. In the above formula $\mu_J$ are the masses of the 4d bifundamental hypermultiplets charged under $(J-1)$th and $J$th gauge groups.
Indeed as pointed out in \cite{Alday:2009fs, Dorey:2011pa}, equation \eqref{eq:BarHiggs2} corresponds to certain degenerate operator insertions in the dual Liouville conformal block, which in turns can be interpreted as surface operator/vortex insertions in the 4d gauge theory.  

The vortex world volume dynamics in this case can be described by a 2d ${\mathcal N}=(2,2)$ gauged linear sigma model with $A_L$ quiver gauge group $\prod_{I=1}^{L} U(N_I)$, whose matter content consists of one adjoint hypermultiplet of mass $\epsilon$ for each $SU(N_I)$; one bi-fundamental hypermultiplet of mass $\frac{\epsilon}{2}$ charged under $SU(N_I)\times SU(N_{I+1})$;  two fundamental hypermultiplets of masses $M_a$ and two anti-fundamental hypermultiplets of masses $\widetilde{M}_a$ charged under $SU(N_1)$. For each gauge factor $U(N_I)$, it also contains an FI parameter $r_I$ and a 2d theta angle $\vartheta_I$, we can combine them into usual complex combination $\hat{\tau}_{I}=i r_I  + \frac{\vartheta_I}{2\pi}$ and also define $\hat{q}_I = e^{2\pi i \hat{\tau}_I}$.

\begin{figure}
\begin{center}
\includegraphics[height=15.5cm, width=12.5cm]{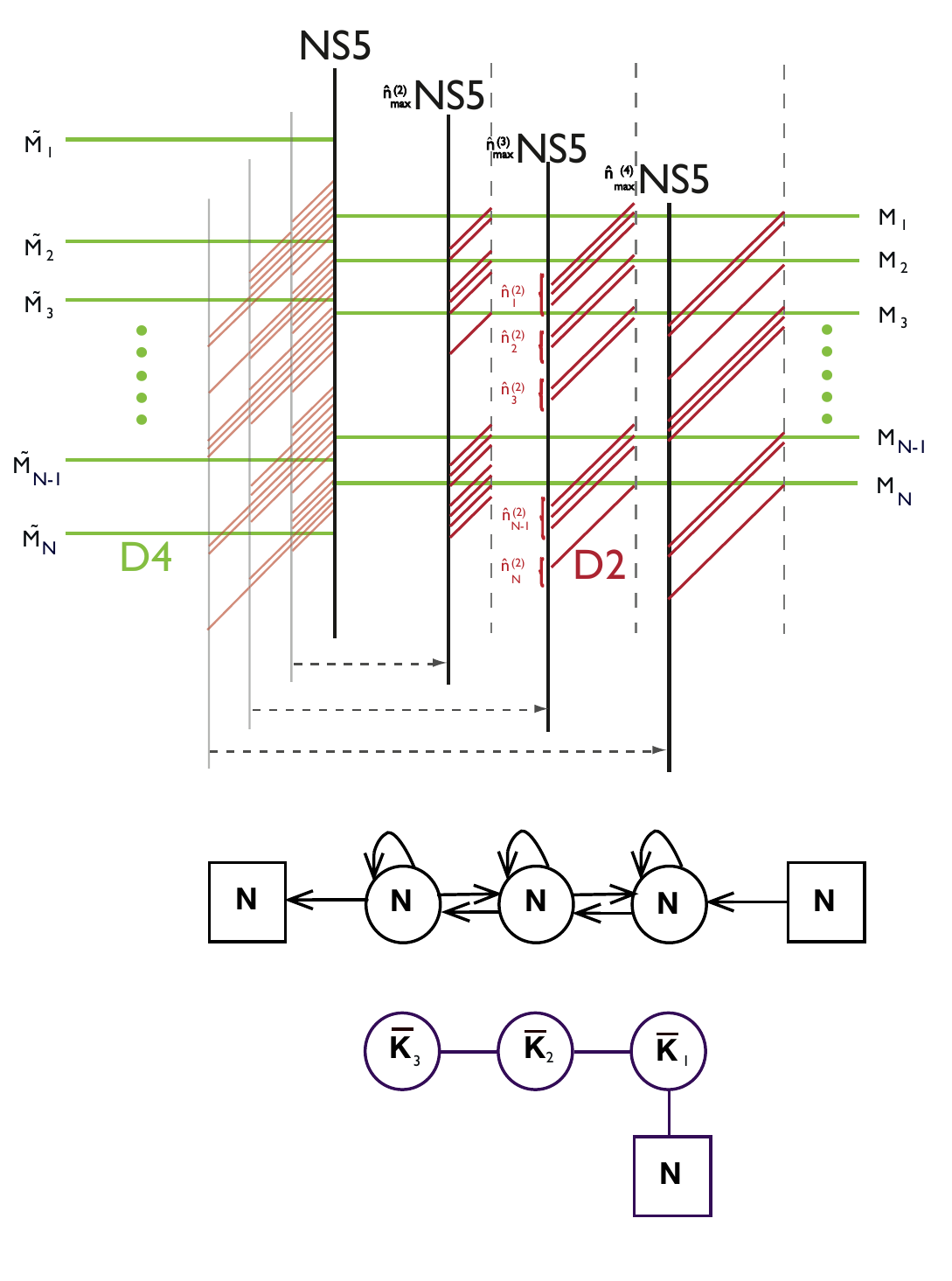}
\caption{The 4d/2d duality for quiver gauge theories. The brane picture on the top of the figure illustrates the Higgs branch of $A_3$ 4d quiver (which is shown in the middle of the figure) theory with $U(N)$ gauge groups at each node. All three different FI parameters are turned on. The D2 branes stretched along the $x^7$ direction show how $a-m$ cycles of the SW curve are quantized along each color direction. Dashed lines on the bottom of the brane picture project D2 branes and NS5 branes down to the $(457)$ space, where the brane construction for the 2d $A_3$ quiver (at the bottom of the figure) can be read off. The NS5 branes at nonzero $x^7$ position appear in stacks with multiplicity $n_{\text{max}}^{(I)}$ in order obey the $s$-rule. The number of colors in each gauge group in the 2d quiver are 
$
\bar{K_3}=K_3\,,\bar{K_2}=K_2+K_3\,, \bar{K_1}=K_1+K_2+K_3\,.
$}
\label{fig:CDHLquiverN}
\end{center}
\end{figure}

From the corresponding D-brane picture for our set up given in \figref{fig:CDHLquiverN}, we can divide the two dimensional color number $N_I$ into
\[
N_I = \sum_{J=I}^L K_J \,, \qquad K_J = \sum\limits_{a=1}^N\hat{n}^{(J)}_a\,,
\]
and we have defined
\begin{equation}
n^{(I)}_a - 1=\sum_{J=I}^L\hat{n}^{(J)}_a\,.
\end{equation}
We can interpret $\hat{n}^{(J)}_a$ as the number of D2 brane stretching between the $a$-th D4 brane and $J$-th NS5 brane, and summing over all the D2 branes ending on different NS5 branes but the same D4 brane, we obtain the vortex number for 4d gauge theory $n_{a}^{(I)}$. Note that given the matter contents, the 2d quiver is no longer conformal, as the vortex number $n^{(I)}_a$ can be chosen arbitrarily. 

The quantum vacua of  2d quiver gauge theory in the vortex world volume can be given by the extrema of the simple one-loop exact twisted superpotential computed in \cite{Witten:1993yc}, which are located at the quantized root of baryonic Higgs branch \eqref{eq:BarHiggs2} and labeled by the flux quanta $\{\hat{n}_a^{(I)}\}$.   
The statement of \cite{Dorey:2011pa, Chen:2011sj}  is that after the change of parameters which will be specified momentarily, 
the on-shell value of the one loop twisted superpotential precisely coincides with the NS lmit Nekrasov partition function (\ref{eq:NSpart}) evaluated at the quantized Higgs root.

\subsubsection{The XXX Spin Chain}
Let us first consider one of the simplest nontrivial 4d quiver theories which leads to an already nontrivial bispectral duality -- the linear quiver with the color-flavor labels $(2,2),(2,0),\dots,(2,0),(2,2)$. The corresponding 2d theory will have the parameter space of vacua which is encoded in the Bethe ansatz equations for such spin chain:
\<
\prod\limits_{a=1}^{2}\frac{\lambda^{(1)}_j-M_a}{\lambda^{(1)}_j-M_a-\kappa_a} \eq q_1\prod_{\substack{i = 1\\ i\neq j}}^{\bar{K}_1}\frac{\lambda^{(1)}_j-\lambda^{(1)}_i-\epsilon}{\lambda^{(1)}_j-\lambda^{(1)}_i+\epsilon}\cdot\prod\limits_{i=1}^{\bar{K}_2}\frac{\lambda^{(1)}_j-\lambda^{(2)}_i-\half\epsilon}{\lambda^{(1)}_j-\lambda^{(2)}_i+\half\epsilon}\,,\nln
1 \eq q_I\prod\limits_{i=1}^{\bar{K}_{I-1}}\frac{\lambda^{(I)}_j-\lambda^{(I-1)}_i-\half\epsilon}{\lambda^{(I)}_j-\lambda^{(I-1)}_i+\half\epsilon}\cdot\prod_{\substack{i = 1\\ i\neq j}}^{\bar{K}_{I}}\frac{\lambda^{(I)}_j-\lambda^{(I)}_i-\epsilon}{\lambda^{(I)}_j-\lambda^{(I)}_i+\epsilon}\cdot\prod\limits_{i=1}^{\bar{K}_{I+1}}\frac{\lambda^{(I)}_j-\lambda^{(I+1)}_i-\half\epsilon}{\lambda^{(I)}_j-\lambda^{(I+1)}_i+\half\epsilon}\,,\nln
1 \eq q_L \prod\limits_{i=1}^{\bar{K}_{L-1}}\frac{\lambda^{(L)}_j-\lambda^{(L-1)}_i-\half\epsilon}{\lambda^{(L)}_j-\lambda^{(L-1)}_i+\half\epsilon}\cdot\prod_{\substack{i = 1\\ i\neq j}}^{\bar{K}_{L}}\frac{\lambda^{(L)}_j-\lambda^{(L)}_i-\epsilon}{\lambda^{L)}_j-\lambda^{(L)}_i+\epsilon}\,,
\label{eq:XXXBAE2}
\>
where $\lambda^{(J)}_i$ are the lowest components of twisted chiral multiplets associated to $J$-th gauge group and $M_a$ and $\kappa_a$ are the 2d mass parameters. We shall specify the relationship between them and the 4d masses $m^{(1)}_a$ and $m^{(L)}_a$ momentarily.
In other words, this set of equations imply that there is an one to one correspondence between the quantum vacua of vortex world volume theory and the eigenstates of $SL(L+1)$ spin chain Hamiltonian.
Moreover in \cite{Dorey:2011pa, Chen:2011sj}, an exact one to one correspondence matching the quantum vacua of the 4d gauge theory in NS limit given by (\ref{eq:BarHiggs2}) and the 2d gauged linear sigma model was further proposed, using the correspondence we can therefore directly identify the 4d quantum vacua with the Hilbert space of $SL(L+1)$ spin chain. This connection turns out to be crucial for us to apply so-called bispectral duality relating the XXX spin chain and the Gaudin model. Here we do not aim to provide a detailed review of \cite{Dorey:2011pa, Chen:2011sj}, but rather only state here the identifications among the parameters of 2d and 4d theories
\[
M_a = m^{(L)}_a - \sfrac{L+2}{2}\epsilon\,,\quad \widetilde{M}_a = m^{(1)}_a + \sum\limits_{J=1}^L\mu_J -\sfrac{L-2}{2}\epsilon\,, \quad \hat{q}_I = (-1)^{N_I+1} q_I\,,
\label{eq:2d4dmasses2}
\]
as required by the exact correspondence. 

By using the quantization condition \eqref{eq:BarHiggs2} and the $U(1)$ constraint $\sum\limits_{a=1}^2 a^{(I)}_a=0$ for each gauge group, we arrive at 
\[
\sum\limits_{a=1}^2 m^{(L)}_a = \sum\limits_{a=1}^2 n^{(I)}_a \epsilon + 2 \sum\limits_{J=I}^{L}\mu_J\,, \quad \mu_L = 0\,.
\label{eq:HiggsandU1}
\]
We can solve the above relationships with respect to the bi-fundamental masses 
\[
\mu_I = \frac{\epsilon}{2}\sum\limits_{a=1}^{2}(n^{(I)}_a-n^{(I+1)}_a)=\frac{\epsilon}{2}K^{(I)}\,, \quad I=1,\dots, L-1\,.
\label{eq:bifundamentalmass}
\]
The equation \eqref{eq:HiggsandU1} for $I=L$ also implies that 
\[
\sum\limits_{a=1}^2 m_a = \sum\limits_{a=1}^2 n^{(L)}_a \epsilon=K^{(L)}+2\,.
\label{eq:spinLpunct}
\]
Analogously to \cite{Bulycheva:2012ct} we can now derive the relationship between the shift parameters $\kappa_a$ of the XXX spin chain, which appear in the in the next section in the first equation of \eqref{eq:XXXBAE} and the gauge theory mass parameters. Using \eqref{eq:BarHiggs2} and \eqref{eq:2d4dmasses2} we get
\[
\epsilon\kappa_a = M_a-\widetilde{M}_a=a^{(1)}_a-m^{(1)}_a + n^{(1)}_a\epsilon-2\epsilon\,,
\label{eq:KappaSpin2}
\]
where we used the fact that \eqref{eq:BarHiggs2} holds for any $I$ and we have chosen $I=1$.
Note that $\kappa_a$ in \eqref{eq:KappaSpin2} is not necessarily an integer.

Let us mention here that the spin chain we are dealing with is anisotropic in two ways. First, as we have already mentioned, there are impurities whose values are given by the twisted masses of the theory. Second, perhaps less expected, spins at different sites of the chain can be different. Indeed, a contribution from each hypermultiplet which contribute to the l.h.s of the first equation in \eqref{eq:XXXBAE2} can be written as 
\begin{equation}
\frac{\lambda_i-M_a}{\lambda_i-M_a-\kappa_a \epsilon}=\frac{\lambda_i-\theta_a+\frac{\epsilon}{2}S_a}{\lambda_i-\theta_a-\frac{\epsilon}{2}S_a}\,,
\end{equation}
where $\theta_a$ and $S_a$ are the impurities and spins of the XXX chain respectively. Remarkably, by tuning the twisted masses $M_a$ and $\widetilde{M}_a$ we can obtain any values of the spin including say, negative or non half-integer.

In what follows we shall make a choice for $m^{(1)}_a$ such that $S_a$ and $\kappa_a$ \eqref{eq:KappaSpin2} become integers. 
\subsubsection{The duality}
\begin{figure}[h!]
\begin{center}
\includegraphics[height=8cm, width=15cm]{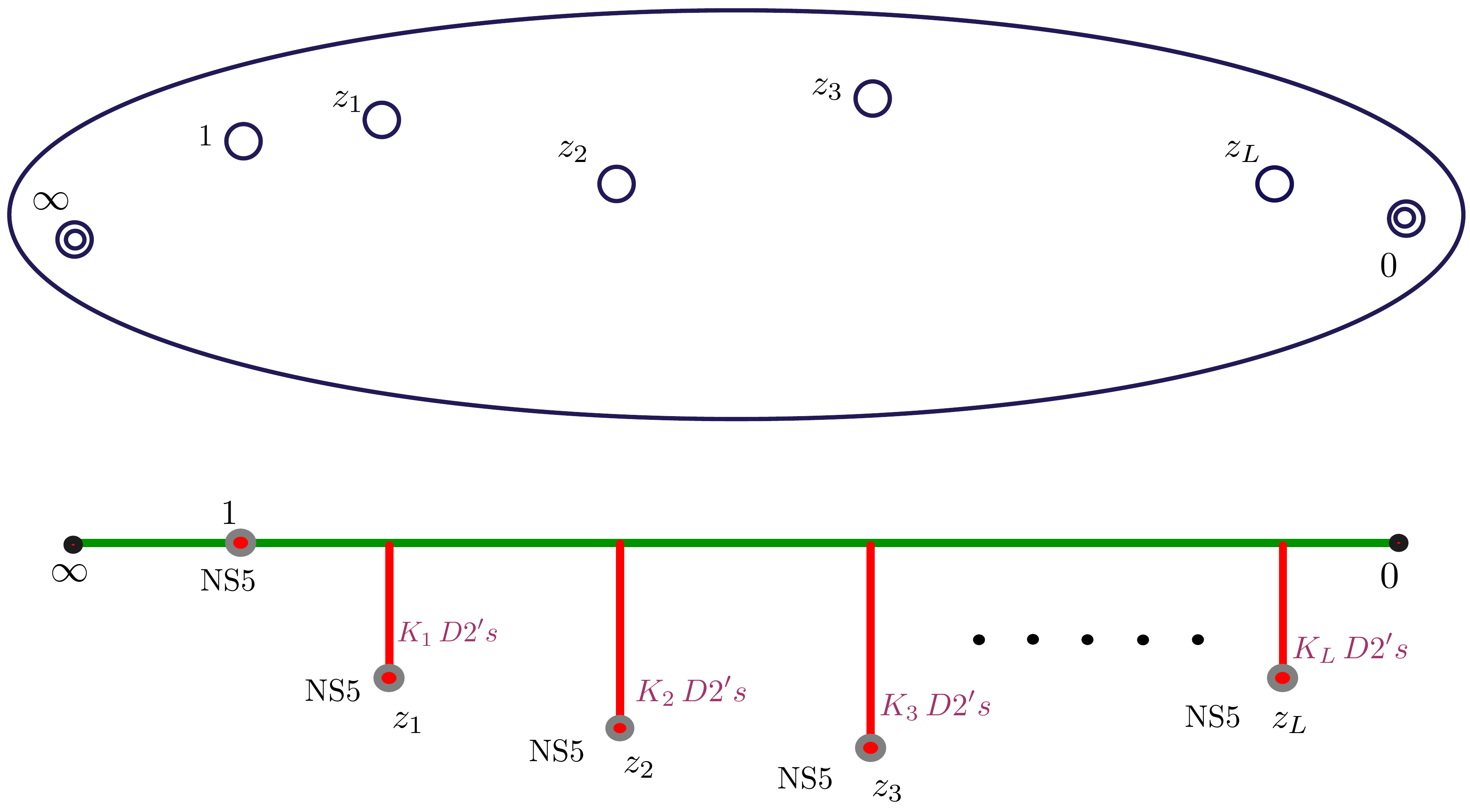}
\caption{Top: two-sphere with $L+3$ punctures. 
Bottom: $67$ slice of the quiver construction in \cite{Chen:2011sj}, see also Fig.~\ref{fig:CDHLquiverN}. Vertical lines correspond to stacks of $D2$ branes. The picture illustrates how to compute the values of Gaudin spins at $z_1,\dots z_L$ by counting the linking numbers of the corresponding NS5 branes.}
\label{fig:CDHLquiver}
\end{center}
\end{figure}

According to the bispectral duality, the XXX $SL(L+1)$ spin chain on two sites whose Bethe ansatz equations \eqref{eq:XXXBAE2} encode the vacua of our 2d/4d gauge theory, can be related to the trigonometric $SL(2)$ Gaudin model on $S^2$ with $L+3$ singularities at points \figref{fig:CDHLquiver} 
\[
z_0=\infty,\quad z_1=1,\quad z_2=q_1,\quad\dots, \quad z_{L+1}=q_1\dots q_L,\quad z_{L+2}=0\,.
\]
The Gaudin model is formulated on a cylinder (a two-sphere with punctures at $0$ and $\infty$) with $L+1$ punctures as shown in \figref{fig:CDHLquiver}.
The Gaudin Bethe equations arise from diagonalization of Gaudin Hamiltonians 
\begin{equation}
\ham_{\text{Gaud}\,i} = \sum_{j\neq i}\frac{\gen{J}_i^a\otimes \gen{J}_{j\,a}}{z_i-z_j}\,,
\end{equation}
acting on $V_1\otimes \dots\otimes V_{L+1}$, where $V_i$ are fundamental representations of $SL(L+1)$. Like XXX spin chain, the Gaudin model can be solved using the Bethe ansatz technique. In the $SL(2)$ case the Gaudin system which is bispectrally dual to the XXX chain \eqref{eq:XXXBAE2} has the following Bethe equations
\begin{equation}
\frac{M_1-M_2-\epsilon}{t_j} +\sum\limits_{I=1}^{L+1}\frac{\nu_I\epsilon}{t_j-z_I} - \sum\limits_{\substack{k = 1\\ k\neq j}}^{\kappa_2}\frac{2\epsilon}{t_j-t_k} = 0\,,
\label{eq:GaudinEqsl2}
\end{equation}
where $\nu_I$ are twice the spins of $SL(2)$ representations sitting at points $z_I$ and $\kappa_2$ is the number of Gaudin Bethe roots. 

Let us now arrange the parameters in $SL(L+1)$ XXX spin chain to facilitate the connection with $SL(2)$ Gaudin model via the bispectral duality. In order to see that other Gaudin spins are given by \eqref{eq:SpinsSL2Quiver} we need to recall that under the bispectral duality, the number of Bethe roots in one system is traded for the spins in the other one. 
At $z_{L+2}=0$ the spin is given by the numerator of the first term in \eqref{eq:GaudinEqsl2} using \eqref{eq:2d4dmasses2} we obtain
\begin{equation}
\nu_{L+2}\epsilon\equiv M_1-M_2-\epsilon = m_1-m_2-\epsilon\,. 
\label{eq:spinl2}
\end{equation}
At the other punctures at $z_1,\dots, z_{L}$ we have
\begin{equation}
\nu_{I} = -N_I+N_{I-1}=K_I\,,\qquad I = 1,\dots, L\,,
\label{eq:spinjI}
\end{equation}
where we used that $N_{L+1}=0$. Therefore we can see that Gaudin spins at $z_1,\dots, z_{L}$ are equal to the linking numbers of the corresponding NS5 branes. 

\subsubsection{Gaudin spins from the spectral curve}
Now we shall discuss an alternative way to see how the Gaudin system arises in the study of the moduli space of vacua of 4d gauge theories -- its spectral curve is nothing but the Seiberg-Witten curve of the corresponding quiver gauge theory.
According to \cite{Witten:1997sc,Gaiotto:2009hg,Nekrasov:2012xe} the curve reads
\begin{equation}
\text{det}(y-\phi(z;z_i))=0\,,
\label{eq:SWcurveHitchin}
\end{equation}
where the Higgs field is 
\begin{equation}
\phi(z;z_i)=\sum\limits_{a=0}^{L+2}\frac{\Phi(z_a)}{z-z_a}\,.
\label{eq:LaxGaudinNP}
\end{equation}
In \eqref{eq:LaxGaudinNP} $\Phi(z_a)$ are the Gaudin Lax matrices.
For the quiver theory we consider in this section the Lax eigenvalues are 
\<
\Phi_{0} \eq m^{(1)}_1-m^{(1)}_2\nln
\Phi_1 \eq m^{(1)}_1+m^{(1)}_2\,,\nln
\Phi_I \eq 2\mu_I\,,\qquad I=2,\dots, L\,,\nln
\Phi_{L+1} \eq  m^{(L)}_1 + m^{(L)}_2\,, \nln
\Phi_{L+2} \eq m^{(L)}_1-m^{(L)}_2\,.
\label{eq:GaudinSpinsNP}
\>
In this paper we are looking at the quantization of the same phase space. Upon quantization of the Gaudin model one arrives to the Bethe equations of the form \eqref{eq:GaudinEqsl2} where the spins $\nu_I$ are precisely the eigenvalues of the Gaudin Lax matrices $\Phi(z)$ (see e.g. \cite{Talalaev:2004qi}).

We now need to compare the above eigenvalues (Gaudin spins) with \eqref{eq:spinl2} and \eqref{eq:spinjI}. One gets
\<
\nu_I \eq K_I\,,\qquad I=2,\dots, L\,,\nln
\nu_{L+1} \eq  K_{L}+2\,, \nln
\nu_{L+2} \eq m^{(L)}_1-m^{(L)}_2-\epsilon\,.
\label{eq:GaudinSpinsOnShell}
\>
The spins sitting at points $z_1,\dots z_L$ \eqref{eq:GaudinSpinsNP} are in complete agreement with those we obtained from the XXX chain via bispectral duality, since, according to \eqref{eq:bifundamentalmass}, $2\mu_l=K_l\epsilon$. However, for $\Phi_{L+1}$ there is a shift as $m_1+m_2=K_L \epsilon+2\epsilon$. It is nevertheless consistent with the result from the Seiberg-Witten curve, since the shift by $2\epsilon$ comes from the shift in the scalar vev $\vev{\phi}=a-\epsilon$. On the XXX side of the bispectral duality this shift corresponds to having $N$ Bethe roots on the last nesting level. Also one may notice a shift in $\Phi_{L+2,a}$. This shift, however, may be absorbed by a different shift of Bethe roots and 2d masses $M_a$ in \eqref{eq:GaudinEqsl2}. Later, when a more generic 2d quiver will be considered, we shall see that the shift will depend nontrivially on the color and flavor labels of the neighbouring nodes of the quiver.

In our approach we cannot compute the spins at $\infty$ and $1$, as masses $m^{(1)}_a$ do not appear in the Higgs branch condition.

\begin{figure}
\begin{center}
\includegraphics[height=2cm, width=4.5cm]{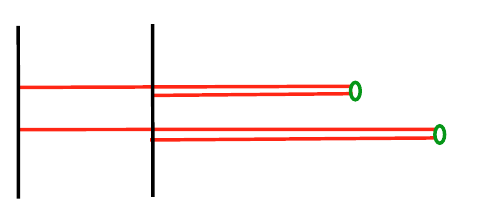}
\caption{Gaudin vacuum state. If D4 branes (green circles) are moved to the left there are no D2 branes (horizontal red lines) in the picture. The configuration in the picture has $N=2$ D2 branes stretched between the NS5 branes (black vertical lines).}
\label{fig:GaudinVac}
\end{center}
\end{figure}

More closely, the state with $N$ D2 branes stretched between the rightmost NS5 brane and the D4 branes such that each D4 brane links one D2 brane corresponds to the vacuum state on the Gaudin side of the duality. Indeed, by looking at the picture \figref{fig:CDHLquiverN} from the right side the above statement is easy to evidence -- let's move all D4 branes further to the right  \figref{fig:GaudinVac}. 
Clearly, in this case the NS5 branes have zero linking numbers with the D4s.

\subsection{Operators in Liouville CFTs and the Gaudin Model}\label{sec:SpherePunct}
Let us now address conformal field theories which arise from 4d quiver gauge theories we are working with in this section using the AGT duality \cite{Alday:2009aq}. The corresponding theories are the the Liouville CFTs on $S^2$ with $L+3$ punctures. 
According to the duality, the equivariant instanton partition function of $\prod_{I=1}^{L} U(2)_I$ quiver gauge theory is equal to the Liouville conformal block which contains $L+3$ primary vertex operators $V_1,\dots, V_{L+3}$ inserted at the following points \footnote{More precisely in \cite{Alday:2009aq}, the authors first considered instanton partition function of linear quiver of $U(2)$s, and it is only after factoring out the $U(1)$ factors we can match the remaining contribution form linear $SU(2)$ quiver with Liouville conformal block. However as subsequently we will introduce vortices/surface operators whose existence depends on the non-trivial first homotopy class of the gauge group, we will keep the $U(1)$ gauge factors.}
\[
\infty,\, 1,\, q_1,\, q_1 q_2,\, \dots,\, q_1 q_2\dots q_{L},\, 0\,,
\label{eq:AllSingularities}
\]
with the following scaling dimensions
\[
\alpha_0(Q-\alpha_0)\,,\quad \mu_0(Q-\mu_0)\,, \dots, \quad \mu_L(Q-\mu_L)\,, \quad \alpha_{L+1}(Q-\alpha_{L+1})\,,
\label{eq:confdimsL}
\]
respectively. These operators are located at the intermediate $s$-channels. In \eqref{eq:confdimsL} $\mu_0$ and $\mu_L$ can be expressed as the following linear combinations of the masses for (anti-) fundamentals of the first and the last gauge groups:
\[
\mu_0=\frac{m^{(1)}_1+m^{(1)}_2}{2}\,,\quad \mu_L=\frac{m^{(L)}_1+m^{(L)}_2}{2}\,,
\]
and 
\[
\alpha_0 = \frac{Q+m^{(1)}_1-m^{(1)}_2}{2}\,,\quad\alpha_{L+1} = \frac{Q+m^{(L)}_1-m^{(L)}_2}{2}\,,\quad \alpha_I = \frac{Q}{2}+a_I \,,
\]
where $a_I = \frac{a^{(I)}_1-a_2^{(I)}}{2}$ is Coulomb branch parameter of $I$-th residual $SU(2)$ gauge group after imposing the traceless condition $a_{1}^{(I)}+a_2^{(I)}=0$.  
Here we also give the relations between the parameters in the gauge theory and the conformal field theory \begin{equation}
\epsilon_1=b \hbar, \quad \epsilon_2=\frac{\hbar}{b}\,,
\end{equation}
where $\epsilon_{1,2}$ are equivariant parameters of the gauge theory,  $\hbar$ is certain mass scale playing the role of Planck constant for the conformal field theory,  while $b$ parameterizes the central charge $c$ of the Virasoro algebra as $c=1+6Q^2$ and $Q=b+\frac{1}{b}$.

Let us look at scaling dimensions of chiral primary operators \eqref{eq:confdimsL} in the WKB NS limit $\epsilon_1\to\infty$ taking into account Higgs branch conditions \eqref{eq:bifundamentalmass} and \eqref{eq:spinLpunct}. We assume that Planck constant $\hbar^2 = \epsilon_1 \epsilon_2$ as $\epsilon_2\to 0$ remains finite, so we get $Q=b =\frac{\epsilon_1}{\hbar}=\frac{\epsilon}{\hbar}$ in this limit. Therefore we obtain
\begin{equation}
\Delta_I = -\epsilon^2 j_I (j_I+1)\,,
\label{eq:sl2Casimir}
\end{equation}
so the scaling dimensions quadratically diverge with $\epsilon$. Spins $j_I$ in the above formula are the following
\<
j_I\,\,\,\,\,\,\,  \eq  K^{(I)}\,,\qquad I = 2,\dots, L+1\,,\nln
j_{L+2} \eq \hat{n}^{(L)}_{1}-\hat{n}^{(L)}_2-1\,,
\label{eq:SpinsSL2Quiver}
\>
where it is assumed that $K^{(L+1)}=0$. Spins at $\infty$ and $1$ formally take negative values in this limit. We can see that the spins sitting at points $z_2,z_3, \dots, z_{L+1}$ correspond to the linking numbers -- total number of D2 branes ending on the $I$'th NS5 brane (see \figref{fig:CDHLquiver}). We can see that \eqref{eq:SpinsSL2Quiver} precisely reproduces Gaudin spins as in \eqref{eq:GaudinSpinsOnShell}. Therefore, one may conclude that the scaling dimensions \eqref{eq:confdimsL} computed via the AGT using the 4d gauge theory data coincide (up to a sign) with the eigenvalues of the (rescaled) quadratic $\alg{sl}_2$ Casimir \eqref{eq:sl2Casimir} on the representations of spins $j_i$ modulo the Higgs branch conditions. 

It is straightforward to generalize the results of this section to higher ranks, viz. to Toda CFT. It is also interesting to systematically derive systems of differential equations which are solved by Toda conformal blocks \cite{BrunoJaume}. Note also that Gaudin model naturally appears in the free field formulation of conformal blocks \cite{Dotsenko1984312,Felder:1988zp}, see also \cite{Gaiotto:2011nm}. 
In \cite{Bonelli:2010gk,Bonelli:2011na} the authors arrived to the Gaudin model by studying matrix models in the NS $\Omega$-background, see also \cite{Bonelli:2011aa} for the investigation to Hitchin systems with wild ramification. 

\subsection{The $\mathfrak{sl}(N)/\mathfrak{sl}(L+1)$ Duality}
Let us now study conformal linear quivers with $SU(N)$ gauge groups at each node. From what we just have considered, when $N=2$, it is a fairly simple generalization. 

\subsubsection{XXX Spin Chain on $N$ sites}
We can write the Bethe Ansatz equations for the anisotropic twisted $\alg{sl}(L+1)$ magnet, represents the ground state equations for the corresponding quiver gauge theory with labels $(\bar{K}_1,N),(\bar{K}_2,0),\dots,(\bar{K}_L,0),(\bar{K}_{L+1},0)$ \cite{Chen:2011sj} :
\<
\prod\limits_{a=1}^{N}\frac{\lambda^{(1)}_j-M_a}{\lambda^{(1)}_j-\widetilde{M}_a} \eq q_1\prod_{\substack{i = 1\\ i\neq j}}^{\bar{K}_1}\frac{\lambda^{(1)}_j-\lambda^{(1)}_i-\epsilon}{\lambda^{(1)}_j-\lambda^{(1)}_i+\epsilon}\cdot\prod\limits_{j=1}^{\bar{K}_2}\frac{\lambda^{(1)}_j-\lambda^{(2)}_i-\half\epsilon}{\lambda^{(1)}_j-\lambda^{(2)}_i+\half\epsilon}\,,\nln
1 \eq q_I\prod\limits_{i=1}^{\bar{K}_{I-1}}\frac{\lambda^{(I)}_j-\lambda^{(I-1)}_i-\half\epsilon}{\lambda^{(I)}_j-\lambda^{(I-1)}_i+\half\epsilon}\cdot\prod_{\substack{i = 1\\ i\neq j}}^{\bar{K}_{I}}\frac{\lambda^{(I)}_j-\lambda^{(I)}_i-\epsilon}{\lambda^{(I)}_j-\lambda^{(I)}_i+\epsilon}\cdot\prod\limits_{i=1}^{\bar{K}_{I+1}}\frac{\lambda^{(I)}_j-\lambda^{(I+1)}_i-\half\epsilon}{\lambda^{(I)}_j-\lambda^{(I+1)}_i+\half\epsilon}\,,\nln
1 \eq q_L \prod\limits_{i=1}^{\bar{K}_{L-1}}\frac{\lambda^{(L)}_j-\lambda^{(L-1)}_i-\half\epsilon}{\lambda^{(L)}_j-\lambda^{(L-1)}_i+\half\epsilon}\cdot\prod_{\substack{i = 1\\ i\neq j}}^{\bar{K}_{L}}\frac{\lambda^{(L)}_j-\lambda^{(L)}_i-\epsilon}{\lambda^{(L)}_j-\lambda^{(L)}_i+\epsilon}\,,
\label{eq:XXXBAE}
\>
where $I=2,\dots, L-1$ in the middle equation,
\[
\bar{K}_l = K_{l+1}+\dots+K_{L+1}\,, \quad l=1,\dots,L\,,
\label{eq:Nlevel}
\]
where the parameters $\nu_k$ are Gaudin spins \eqref{eq:GaudinSpins} and $q_I=z_{I+1}/z_I$. By making the following transformation
\<
\lambda^{(J)}_i=\epsilon\, t^{(J)}_i - \sfrac{J-1}{2}\epsilon\,,\quad J = 1,\dots,L\,,
\>
and
\[
M_a = -\mathcal{M}_a-(a-1)\epsilon\,,\quad \widetilde{M}_a = -\mathcal{M}_a-(a-1)\epsilon-\kappa_a \epsilon\,,
\label{eq:MassvsAnisotr}
\]
where $a=1,\dots,N$, we arrive to the $\alg{sl}(L+1)$ Bethe Ansatz equations as they are presented in \cite{MR2409414}. 

\subsubsection{The $\mathfrak{sl}(N)$ Gaudin Model}
The $\mathfrak{sl}(N)$ Gaudin model on $S^2$ with $L+3$ marked points $z_j$ \eqref{eq:AllSingularities} has the following Yang-Yang function 
\<
Y(t^{(1)},\dots,t^{(N-1)})\eq \sum\limits_{j=1}^{\bar{\kappa}_i}\sum\limits_{i=1}^{N}(\mathcal{M}_i-\mathcal{M}_{i+1}-\epsilon)\log\left(t^{(i)}_j\right) +\sum\limits_{i=1}^{N}\sum\limits_{j=1}^{\bar{\kappa}_i}\sum\limits_{a=1}^{L+1} \nu_a \log\left(t^{(i)}_j-z_a\right)\nl
-2\sum\limits_{i=1}^{N-1}\sum\limits_{\substack{k = 1\\ k\neq j}}^{\bar{\kappa}_i}\log\left(t^{(i)}_j-t^{(i)}_k\right)
+\sum\limits_{i=1}^{N-2}\sum\limits_{j=1}^{\bar{\kappa}_{i}}\sum\limits_{\substack{
k = 1\\ k\neq j}}^{\bar{\kappa}_{i+1}}\log\left(t^{(i)}_j-t^{(i+1)}_k\right)\,,
\label{eq:YYGaudin}
\>
The corresponding Bethe equations are obtained from \eqref{eq:YYGaudin} by differentiating w.r.t. $t^{(i)}_j$'s read \cite{MR2409414}
\<
\frac{\nu^{(1)}_0}{t^{(1)}_j} +\sum\limits_{a=1}^{L+1}\frac{\nu^{(1)}_a}{t^{(1)}_j-z_a} - \sum\limits_{\substack{k = 1\\ k\neq j}}^{\bar{\kappa}_1}\frac{2}{t^{(1)}_j-t^{(1)}_k} + \sum\limits_{k = 1}^{\bar{\kappa}_2}\frac{1}{t^{(1)}_j-t^{(2)}_k} \eq 0\,,\nln
\frac{\nu^{(i)}_0}{t^{(i)}_j} - \sum\limits_{\substack{k = 1\\ k\neq j}}^{\bar{\kappa}_i}\frac{2}{t^{(i)}_j-t^{(i)}_k} + \sum\limits_{k = 1}^{\bar{\kappa}_{i-1}}\frac{1}{t^{(i)}_j-t^{(i-1)}_k}  + \sum\limits_{k = 1}^{\bar{\kappa}_{i+1}}\frac{1}{t^{(i)}_j-t^{(i+1)}_k} \eq 0\,, \nln
\frac{\nu^{(N-1)}_0}{t^{(N-1)}_j} - \sum\limits_{\substack{k = 1\\ k\neq j}}^{\bar{\kappa}_{N-1}}\frac{2}{t^{(N-1)}_j-t^{(N-1)}_k} + \sum\limits_{k = 1}^{\bar{\kappa}_{N-2}}\frac{1}{t^{(N-1)}_j-t^{(N-2)}_k} \eq 0\,,
\label{eq:BAEGaudinslN}
\>
where $i=2,\dots,N-2$ and $j=1,\dots,\bar{\kappa}_i$. There are $\sum\limits_{i=1}^{N}\bar{\kappa}_i$ Bethe equations in \eqref{eq:BAEGaudinslN}. The following representation of the parameters of the Gaudin model is convenient to study the bispectral duality
\[
\bar{\kappa}_i=\kappa_{i+1}+\dots+\kappa_{N}\,,
\label{eq:Klevel}
\]
It will soon become clear why such parameterization is chosen. Also Gaudin spins are parameterized as
\[
\nu^{(i)}_0 = \mathcal{M}_i-\mathcal{M}_{i+1}-\epsilon\,, 
\]
and 
\[
\nu^{(1)}_j=\nu_j\,,\quad j =1,\dots, L+1\,,
\label{eq:GaudinSpins}
\]
are the spins of $\mathfrak{sl}(N)$ representations at punctures $z_j$. More generally we can write Gaudin Bethe equations as 
\[
\sum\limits_{a=0}^{L+2}\frac{\vev{\nu_I,\gen{e}_a}}{t^{(I)}_j-z_a} =
\sum\limits_{\substack{k = 1\\ k\neq j}}^{\bar{\kappa}_I}\frac{\gen{C}_{II}}{t^{(I)}_j-t^{(I)}_k} + \sum\limits_{\substack{K = 1\\ K\neq I}}^{N-1}\sum\limits_{k = 1}^{\bar{\kappa}_{K}}\frac{\gen{C}_{IK}}{t^{(I)}_j-t^{(K)}_k}\,,
\label{eq:BAEGaudinslNGen}
\]
where $I=1,\dots, N-1$, $j=1,\dots,\kappa_I$, and $\gen{C}_{IK}$ are components of the Cartan matrix of the Lie algebra.

\subsubsection{Gaudin Spins from the SW curve}
From the SW solution we get the following eigenvalues of the Lax matrices
\<
\Phi_{0,a} \eq \{m^{(1)}_a-m^{(1)}_{a+1}\}\nln
\Phi_1 \eq \sum_{a=1}^N m^{(1)}_a \,,\nln
\Phi_I \eq N\mu_I\,,\qquad I=2,\dots, L\,,\nln
\Phi_{L+1} \eq \sum_{a=1}^N m^{(L)}_a \,, \nln
\Phi_{L+2,a} \eq \{m^{(L)}_a-m^{(L)}_{a+1}\}\,.
\label{eq:GaudinSpinsNPGen}
\>
The corresponding Gaudin spins are
\<
\nu_I \eq K_I\,,\qquad I=2,\dots, L\,,\nln
\nu_{L+1} \eq  K_{L}+N\,, \nln
\nu_{L+2,a} \eq \{m^{(L)}_a-m^{(L)}_{a+1}-\epsilon\}\,.
\label{eq:GaudinSpinsOnShellGenN}
\>
Again, we find a nice agreement, since, analogously to \eqref{eq:bifundamentalmass} we get
\begin{equation}
\mu_I = \frac{K_I}{N}\,,
\end{equation}
and the shift by $N$ in the value of $\nu_{L+1}$ in \eqref{eq:GaudinSpinsOnShellGenN} is due to the shift of the scalar vev of the four dimensional theory in the $\Omega$-background.

\subsubsection{The Duality}
Analogously to \eqref{eq:KappaSpin2} we obtain the following condition
\begin{equation}
\kappa_a = n^{(1)}_a-2 = \sum\limits_{I=1}^{L}\hat{n}^{(I)}_a-1\,.
\label{eq:LevelMatchingGen}
\end{equation}
Therefore we have the following expression for the total number of Gaudin Bethe roots
\begin{equation}
\sum\limits_{I=1}^{L} K_I +(K_{L+1}+N) - N= \sum\limits_{I=1}^{L+1} K_I\,,
\label{eq:GaudinBetheRoots}
\end{equation}
where the shift at the $(L+1)$st position is due to the choice of the Gaudin vacuum \figref{fig:GaudinVac}.

According to the MTV paper \cite{MR2409414} Bethe ansatz equations of the XXX spin chain \eqref{eq:XXXBAE} and of the Gaudin model \eqref{eq:BAEGaudinslN} have isomorphic spaces of orbits of solutions provided that the following level matching condition on Bethe roots of both models holds
\[
\sum\limits_{a=1}^N \kappa_a+N = \sum\limits_{J=1}^{L+1} K_J\,,
\label{eq:levelmatching}
\]
where $\kappa_a$ and $\nu_J$ are anisotropies of the XXX chain and spins of the Gaudin system respectively and are related to the numbers of Bethe roots at each nesting level via \eqref{eq:Klevel,eq:Nlevel}. We see from \eqref{eq:GaudinBetheRoots} that \eqref{eq:levelmatching} is indeed true. The condition simply equates different ways of counting of the total number of the D2 branes (see \figref{fig:CDHLquiverN}). 

The precise isomorphism between Gaudin and XXX Bethe roots can be worked out explicitly. It is, however, not convenient to do straightforwardly due to the presence of high degree polynomial equations. Instead one may compare certain rational functions of rapidities and massive parameters which can be deduced by differentiating Yang-Yang functions (which according to the NS duality coincide with the twisted superpotentials) of both models \cite{Gaiotto:2013bwa}. Schematically they read as follows
\begin{equation}
p^m_i=\frac{\partial \mathcal{W}(\lambda,m_i,z_a)}{\partial m_i}\,,\qquad p^z_a=\frac{\partial \mathcal{W}(\lambda,m_i,z_a)}{\partial z_a}\,.
\label{eq:ConjugateMomenta}
\end{equation}
Remarkably those quantities play the role of conjugate momenta on the moduli space of supersymmetric vacua of the 2d gauge theory which gives rise to the XXX chain. Upon the bispectral duality tuple of momenta $(p^m_i,p_a^z)$ \eqref{eq:ConjugateMomenta} for the XXX are to be identified with analogous set of momenta $(\tilde p_a^z, \tilde p^m_i)$ for the Gaudin system. Note that the Gaudin momenta are interchanged.

Thus the bispectral duality maps the solutions of Bethe ansatz equations of $\mathfrak{sl}(L+1)$ XXX chain on $N$ sites and $\mathfrak{sl}(N)$ trigonometric Gaudin model on $L+1$ sites. Upon the duality spins on one model are mapped onto the number of Bethe roots of the other and vice versa (see \tabref{tab:BispectralDualityTable}).
\begin{table}[h!]
\centering
\begin{tabular}{ | c | c | c | c | }
 \hline
XXX chain  & Trigonometric Gaudin model \\ 
\hline\hline
$\mathfrak{sl}(L+1)$ symmetry & $L+1$ punctures on $S^2-\{0,\infty\}$ \\
\hline
Chain length $N$ & $\mathfrak{sl}(N)$ symmetry \\
\hline
Twist parameters $q_I$  & Impurities (positions of punctures) \\
\hline
Impurities $m_a$ & Twists (spins at $0$ and $\infty$) \\
\hline
Spins $\kappa_a$  &  Number of Bethe roots $\bar{\kappa}_a$  \\
\hline
Number of Bethe roots $\bar{K}_I$ & Spins $K_I$ at degenerate punctures  \\ 
\hline
\end{tabular}
\caption{The bispectral duality in the details.}
\label{tab:BispectralDualityTable}
\end{table}

\subsection{In Full Generality}
Let us now consider the most general conformal linear $A_L$ quiver in four dimensions such that its color and flavor labels obey \eqref{eq:betafuncitonzero}, so beta-functions vanish for each coupling. Nekrasov and Pestun's result \cite{Nekrasov:2012xe} provides us with the eigenvalues of the Lax matrices of the Hitchin system on $S^2$ with $L+3$ punctures for such a quiver. The result can be obtained systematically by expanding function $z(y)$ from \eqref{eq:LaxGaudinNP} of the Gaudin model near $x=\infty$ and fishing out the coefficients in front of the leading terms. Matrices corresponding to the singularities at $0$ and $\infty$ will have maximal ranks, whereas those at other punctures have rank one (so-called \textit{full} vs.\!\! \textit{simple} punctures). The eigenvalues of $\Phi_I(z_i)$ \eqref{eq:LaxGaudinNP} at $z_1,\dots,z_L+1$ are given as linear combinations of the fundamental and the bifundamental masses. In what follows we shall explain how these formulae arise from the solution we have described above (see \eqref{eq:GaudinSpinsNPGen}) for the quiver with $N$ colors at each node.

\begin{figure}
\begin{center}
\includegraphics[height=6cm, width=14cm]{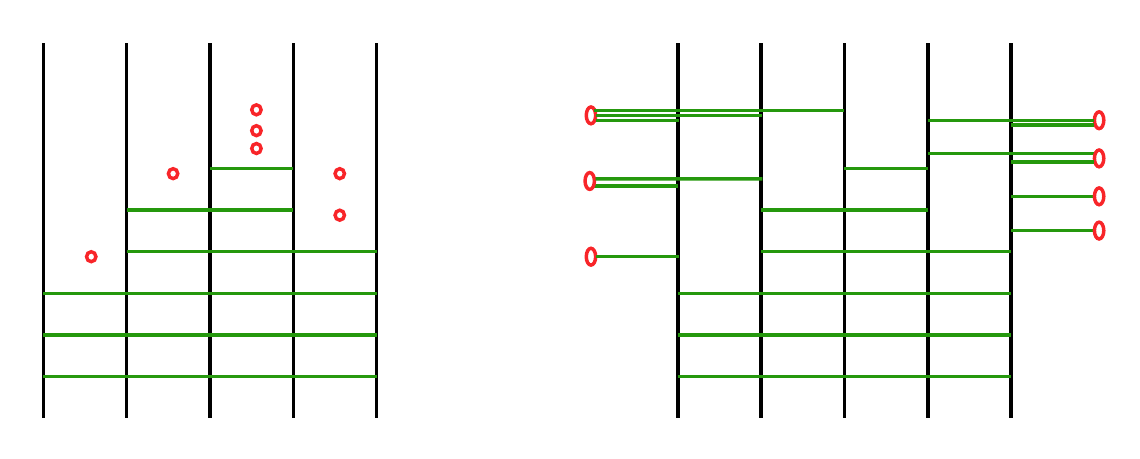}
\caption{Linear conformal $A_4$ quiver with labels $(3,1),(5,1),(6,3),(4,2)$ in two equivalent configurations. The construction on the right can be obtained from the one on the left by passing flavor D6 branes (red ovals) through NS5 branes (vertical black lines). During this process Hanany-Witten phenomena occur and the corresponding D4 branes (horizontal green lines) are created. These new fourbranes are locked up in the $45$ plane to the values of the corresponding mass $m_a^{(J)}$. On the right the balanced distribution of sixbranes along the quiver is shown.}
\label{fig:NPHW}
\end{center}
\end{figure}

\begin{figure}
\begin{center}
\includegraphics[height=4.5cm, width=14cm]{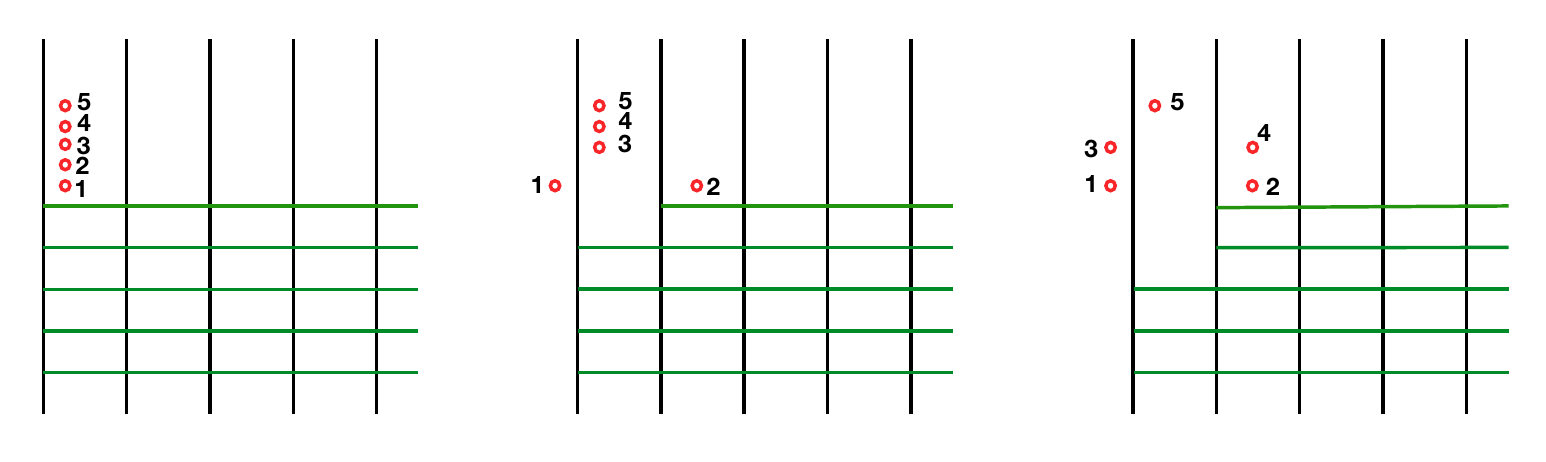}
\caption{Example of Higgsings and Hanany-Witten moves on a part of the linear quiver. One starts from $(5,5),(5,0),(5,0),(5,0),\dots$ quiver and gets respectively $(4,3),(5,1),(5,0),(5,0),\dots$ and $(3,1),(5,2),(5,0),(5,0),\dots$ quivers.}
\label{fig:HWHiggs}
\end{center}
\end{figure}

Let us look at the brane picture for such a quiver (see \figref{fig:NPHW}) with labels 
\begin{equation}
(3,1),(5,1),(6,3),(4,2), 
\label{eq:Ex1}
\end{equation}
One can show that such a quiver can be obtained from another quiver, which we have already discussed above, by a certain sequence of Higgsings and Hanany-Witten moves. We shall not spend much time on reviewing the idea, the reader is welcome to take it from \cite{Gaiotto:2013bwa}. Briefly, the idea is to align flavor branes with some color branes (Higgsing), e.g. $a^{(I)}_a=m^{(I)}_1=m^{(I)}_2+\epsilon$, then, by removing a part of the D4 brane between the flavor branes off to infinity, and further Hanany-Witten moves of the flavor branes, one arrives to a different quiver. See the example in \figref{fig:HWHiggs}. 
Note that in \cite{Gaiotto:2013bwa} the procedure was implemented for triangular quivers (albeit in the T-dual IIB formalism). Evidently the procedure can be generalized to quivers of the form $(N,N),(N,0),\dots,(N,0),(N,N)$ by Higgsing it from the left and from the right simultaneously. Let us call such quivers $Q[U(N)]$, where $U(N)$ stands for the flavor symmetry on the edge nodes \footnote{In fact there are two copies of $U(N)$.}. Each step will eliminate some color and flavor branes from the left and from the right of the skeleton until the two destructive forces meet at some node $1\leq I_\ast\leq L$, e.g. $I_\ast = 3$ in \figref{fig:NPHW}. So if one starts with $Q[U(6)]$, after Higgsings and Hanany-Witten moves one arrives to the new quiver $Q[U(6)]^{\rho_-\,\rho_+}$, where $\rho_{\pm}$ are partitions which are defined by flavor labels $M_I$ of the quiver.
We can dial these partitions by Young tableaux whose rows are 
\begin{equation}
r_-^I = \sum_{J=I}^{I_\ast-1}(M_J+M^{-}_{I_\ast})\,,\quad r_-^{I_\ast}= M^{-}_{I_\ast}\,,\qquad r_+^I = \sum_{J=I_\ast+1}^{I} (M_J+M^{+}_{I_\ast})\,,\quad r_+^{I_\ast}=M^{+}_{I_\ast}\,,
\label{eq:BalancedMotion}
\end{equation}
where 
\begin{equation}
M^{\pm}_{I_\ast} = N_{I_\ast}-N_{I_\ast\pm 1}\,.
\end{equation}
It is easy to see that 
\begin{equation}
M^{+}_{I_\ast}+M^{-}_{I_\ast}=2N_{I_\ast}-N_{I_\ast- 1}-N_{I_\ast+ 1}=M_{I_\ast}\,,
\end{equation}
due to the zero beta-function condition on the $I_\ast$ node. Also by construction it is clear the the sizes of tableaux $\rho_-$ and $\rho_+$ are the same
\begin{equation}
\sum_{J=I}^{I_\ast}r_-^I=\sum_{J=I}^{I_\ast-1}I M_J + I_\ast M^{-}_{I_\ast}=I_\ast M^{-}_{I_\ast}+\sum_{J=I}^{I_\ast+1}I M_J =\sum_{J=I_\ast}^{L}r_+^I= N_{I_\ast}\,.
\end{equation}
We can see that the number of $D4$ branes in \figref{fig:HWHiggs} for between each pair of adjacent NS5 branes is equal to six, same as in $Q[U(6)]$ theory. Also $N_{I_\ast}=6$. In \cite{Gaiotto:2009hg} such a distribution of flavor branes, when there are $r_-^1$ flavor branes to the left and $r_{+}^{L}$ branes to the right of the boundary of the segment along $x^6$ where NS5 branes are located at, is called ``balanced'' distribution, which results in even distribution of the $D4$ branes along the segment. From the right picture in \figref{fig:HWHiggs} we can see that some D4 branes appear in the locked position

The Seiberg-Witten solution for conformal (and also asymptotically free) linear quivers and its connection to the Hitchin system has been discussed in the literature in the great detail \cite{Witten:1997sc,Gaiotto:2009hg,Nekrasov:2012xe}. In particular, the SW curve in the form \eqref{eq:SWcurveHitchin} coincides with the spectral curve of the Hitchin system, where $\phi(z)$ is the Higgs field. Residues of the Higgs field at $\infty, 1, z_1,\dots z_{L}, 0$ are given by Lax matrices of the Hitchin system (the Gaudin model in a cylinder) \eqref{eq:LaxGaudinNP}.

\begin{figure}
\begin{center}
\includegraphics[height=5cm, width=14cm]{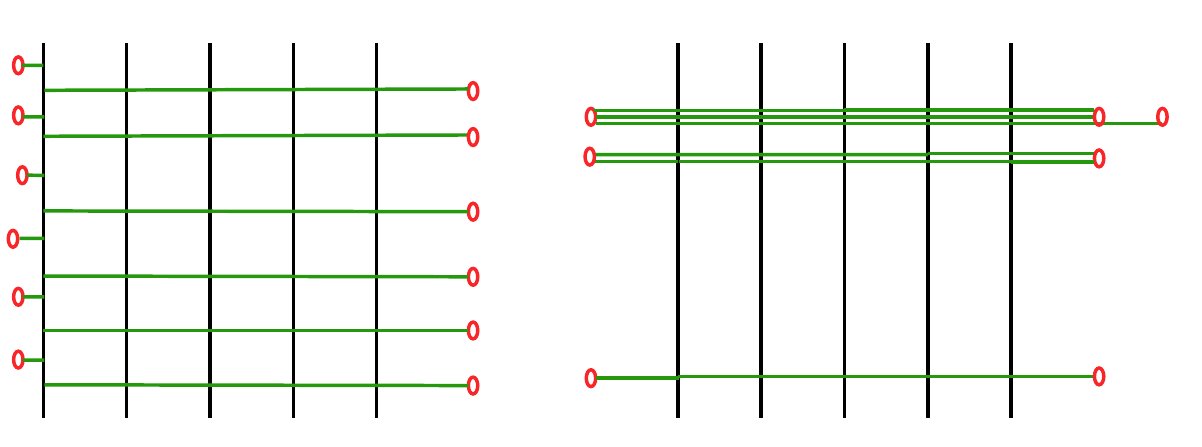}
\caption{Higgsing the conformal quiver. Left: Higgsed $Q[U(6)]$ quiver; right: fully Higgsed $Q[U(6)]^{\rho_- \rho_+}$ quiver with $\rho_- =(4,3,1)$ and $\rho_+=(5,3)$ (lengths of rows in partitions are given).}
\label{fig:NPHWHiggs}
\end{center}
\end{figure}

\begin{figure}
\begin{center}
\includegraphics[height=3.3cm, width=3.6cm]{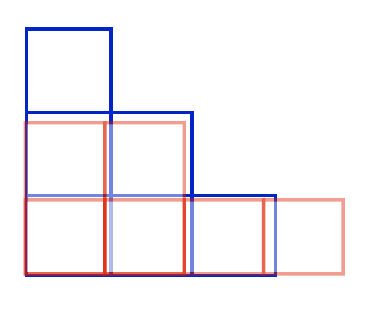}
\caption{Intersecting Young tableux of $\rho_-$ (in blue) with $\rho_+$ (in red).}
\label{fig:YTint}
\end{center}
\end{figure}

We now return to the main subject of this section -- the bispectral tGaudin/XXX duality. 
In order to follow the logic which we have outlined in the beginning of this section (see \figref{fig:flowchartsphere}), we need to be able to put the 4d theory in the Higgs phase and obtain the 2d GLSM whose ground state equations will lead to the XXX Bethe equations.
In \figref{fig:NPHWHiggs} we show two quivers in the Higgs phase.
We can see that in order to Higgs a generic quiver of the $Q[U(N)]^{\rho_- \rho_+}$ type (say, the one from \figref{fig:HWHiggs}) some fourbranes need to be locked together (to be located at the same position in the $(45)$ plane). It is also quite easy to see what the corresponding degeneracies (numbers of D4 branes in each stack) should be once the partitions $\rho_{\pm}$ are known. Indeed, \figref{fig:HWHiggs} suggests that the number of fourbranes ending on each flavor brane, in the configuration when all D6 branes are moved off to the boundary, precisely gives heights of columns of Young tableaux $\rho_{\pm}$. As is argued in \cite{Witten:1997sc,Gaiotto:2009hg} these numbers also specify degeneracies of the Lax matrix eigenvalues at zero and infinity. In the example given in \figref{fig:HWHiggs} we have for $\phi_0$ the following eigenvalues $m^{(3)}_1$ with degeneracy three, $m^{(2)}_1$ with degeneracy two and $m^{(1)}_1$ nondegenerate. Similarly, at $z=0$ for $\phi_{L+2}$ we have two doubly degenerate eigenvalues $m^{(3)}_{2,3}$ and two nondegenerate ones $m^{(4)}_{1,2}$. In order to fully Higgs the quiver we need to ``intersect'' tableaux for $\rho_+$ and $\rho_-$ as shown in \figref{fig:YTint}.
The degeneracy of D4 branes is defined as follows. One takes the tableau with the smallest number of columns, then all the boxes to the right of the last column of the ``narrowest'' tableau have to be placed on top of this tableau in order to match the hight of the the highest column at each position. We shall call such partition $\rho_{\cap}$. Thus, in our example $\rho_-$ has three columns and $\rho_+$ has four, so the last column of $\rho_+$, consisting of a single block, is placed on top of the first column of $\rho_+$ in order to match the same column for $\rho_-$. On the brane language it results in the situation depicted on the right picture in \figref{fig:NPHWHiggs}, there are two branes at the same location in $(45)$ plane on the right boundary e.g. $m^{(4)}_1=m^{(3)}_1$. In this case $\rho_\cap$ coincides with $\rho_-$, however, for more generic configurations it may be different.

Clearly the above tableaux intersection argument constraints the parameter space of vacua of the 4d theory for the latter to be Higgsed. For instance, there is only a single sixbrane on one of the boundaries, so the corresponding tableau is a height-$N$ column, all other branes on the opposite boundary have to be locked up together. Therefore, for each set of color and flavor labels of a 4d quiver there is a (generally) constrained space of parameters which allows us to study the theory in the Higgs branch.

\begin{figure}
\begin{center}
\includegraphics[height=5cm, width=5.5cm]{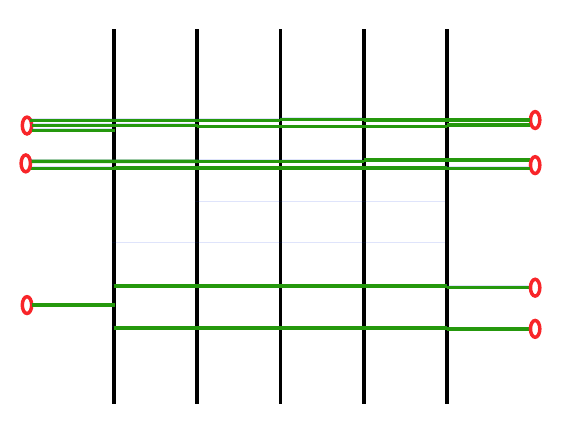}
\caption{Higgsing the fundamental matter (four flavor sixbranes located to the right of $I_\ast=3$) of quiver (\ref{eq:Ex1}). While moving into the Higgs branch of the theory all NS5 branes but the leftmost one will be moved along the $x^7$ direction.}
\label{fig:NPHiggsAll}
\end{center}
\end{figure}

However, one does not have to Higgs all the flavors, as shown on the right in \figref{fig:NPHWHiggs}. We may only Higgs those sixbranes which are located the right of the $I_\ast$ node and all bifundamentals in-between. Therefore, all gauge and matter fields in the theory will become color-flavor locked, except the junction on the leftmost NS5 brane (see \figref{fig:NPHiggsAll}).
In this case $\rho_\cap=\rho_+$ is always true for any partition $\rho_+$. 

Finally, the configuration depicted on the right picture of \figref{fig:NPHWHiggs} can be viewed as the Higgsed $Q[U(N)]$ quiver, where the fourbranes are collected in stacks according to $\rho_\cap$. Indeed, in order to see this one needs to move all the sixbranes to left (right) infinity in the $x^6$ direction. Therefore we have just reduced the problem to the one we solved above for the $Q[U(N)]$ quiver. Indeed, we can simply impose the constraints on Coulomb branch parameters, fundamental and bifundamental masses, such that the partition $\rho_\cap$ is formed as described above, on the Higgs branch conditions \eqref{eq:BarHiggs2}. Then the corresponding XXX spin chain and tGaudin Bethe equations can be read off easily. The dual pair of equations will be represented by \eqref{eq:BAEGaudinslN} and \eqref{eq:XXXBAE} with the proper constraints on masses being imposed. 

Now let us assume that all NS5 branes can be Higgsed, that happens when all D3 branes to the left and to the right of the brane 
picture can be aligned with those on the right. In particular, for the $Q[U(N)]$ quiver it happens when $m^{(1)}_a-m^{(L)}_a=\epsilon$. We can now look at the 2d projection of the brane construction. Interestingly enough, the Higgsings and Hanany-Witten moves, which we have described earlier in this section, allow us to study XXX chains in a more generic representation of $SL(L+1)$ then in merely the fundamental one. Thus for a generic conformal quiver the representation is the following (see \cite{Gaiotto:2013bwa} for more details)
\begin{equation}
\mathcal{R} = \bigoplus_{i=1}^{\text{m}_1}\textbf{(L+1)} \oplus \bigoplus_{i=1}^{\text{m}_2}\Lambda^2\textbf{(L+1)} \oplus \dots \oplus \bigoplus_{i=1}^{\text{m}_{L}}\Lambda^{L}\textbf{(L+1)}\,,
\label{eq:RepSpinChain}
\end{equation}
i.e. $\text{m}_1$ copies of the fundamental representation with impurity parameters $m_a^{(1)}$ , $\text{m}_2$ copies of the second antisymmetric power of the fundamental representation of $SL(L+1)$ with impurity parameters $\text{m}_a^{(2)}$, etc. Following the Higgsing procedure the parameters satisfy $\sum_{a=1}^{N_\ast} a\,\text{m}_a=N_\ast$. The easiest way to obtain a chain in representation $\mathcal{R}$ is to implement the constraints on the Coulomb branch parameters and masses (see \figref{fig:HWHiggs}) on the XXX Bethe equations \eqref{eq:XXXBAE}. For example, if $\lambda^{(1)}=m^{(1)}_1=m^{(1)}_2-\kappa_1\epsilon$ then one of the terms in the left hand side of the first equation in \eqref{eq:XXXBAE} is destroyed, and in the second equation, the term containing $\lambda^{(1)}=m^{(1)}_1$ can be moved to its l.h.s. A transformation of this kind creates the second antisymmetric power of the fundamental representation in \eqref{eq:RepSpinChain}. As a consequence on the Gaudin side of the duality at $z=0$ some $\nu^{(0)}_a$s will have fixed values $(\kappa_a-1)\epsilon$.
  
Therefore we have proved the bispectral duality between the XXX chain and Gaudin model as conjectured in \cite{2006math......4048M} and generalized it the spin chains transforming under more complicated representations like \eqref{eq:RepSpinChain}.

\section{Toda and WZNW Conformal Blocks on a Torus with Punctures}\label{sec:Torus}
In this section we study 6d $(0,2)$ theory compactifications on a four-manifold $X_4$ in presence of two types of defects. Defects of the  first type {\bf (A)} wrap the punctured Riemann surface and fill a two-dimensional subspace inside $X_4$, which is the worldvolume of  a sought 4d gauge theory (see \tabref{tab:defects}). 
\begin{table}[h!]
\centering
\begin{tabular}{ | c | c | c | c | }
 \hline
defect  & $X_4$ & $C_2$ & effect to theory on $C_2$ \\ \hline
(A)  &  2 & 2 & change Toda to WZNW \\
(B)  &  2 & 0 & degeneration in Toda field \\ 
  \hline
\end{tabular}
\caption{Two Different Classes of Surface Operators}
\label{tab:defects}
\end{table}
The two-dimensional dual conformal field theory for such configuration was conjectured to be the affine $SL(N)$ WZNW model in \cite{Alday:2010vg}. Later in \cite{Tan:2013tq} the conjecture was proven using the M-theory. For the high rank $SU(N)$ gauge theories, there is however another closely related two dimensional conformal field theory, namely the $A_{N-1}$ conformal Toda field theory constrained by $\mathcal{W}_N$  symmetry algebra, whose conformal block was conjectured in \cite{Wyllard:2009hg} to coincide with the Nekrasov partition function of four dimensional linear $SU(N)$ quiver gauge theory. Furthermore it was conjectured in \cite{Kozcaz:2010af} that, the degenerated vertex operator insertion with momentum labeled by fundamental weight of $SU(N)$ corresponds to the introduction of surface operator insertions charged under the Levi subgroup of $SU(N)$; and from the 6d perspective this class of surface operators terminate as punctures on the corresponding Riemann surface. We call this class of surface operators  defect {\bf (B)} and they will motivate our later consideration of ${\mathcal W}_N$ conformal block with degenerate vertex operator insertions. We shall in particular focus on rank two theories thereby providing a generalizations to the results of \cite{Hikida:2007tq}. 

Despite different 6d origins of defects {\bf (A)} and {\bf (B)}, from the perspective of the 4d gauge theory on $X_4$ obtained by compactifying M5 branes on Riemann surface $C_2$, with appropriate charge assignments under the gauge subgroup,
these two classes of surface operators should generate exactly the same singularities in the gauge theory worldvolume, and resulting in the same instanton partition function. Through the AGT correspondence, we therefore expect agreement between the conformal block with degenerated insertions in Toda field theory, and the affine conformal block with additional insertion of group element $\mathcal{K}$ described in \cite{Alday:2010vg, affineSLN}. In fact for the rank one case, the effect of $\mathcal{K}$ in affine $SL(2)$ conformal block is unimportant for the matching with instanton partition function, and the relevant conformal blocks for both Liouville and affine $SL(2)$ conformal field theories coincide exactly as shown in \cite{H3+1} for arbitrary genus $g$ up to a coordinate transformation (``twisting factor") from Sklyanian's separation of variables.

For $N=3$, without introducing the ${\mathcal K}$ factor, Ribault in \cite{Ribault:2008si} argued the matching between two conformal blocks on the sphere can hold true only at critical level $k\to -N$ of affine algebra, by identifying third-level BPZ equation and three-current KZ equation obeyed by corresponding correlation functions. The critical level limit corresponds to limit $b\to \infty$ in the Toda theory, or equivalently NS limit $\epsilon_2\to 0$. 
However from the prior discussion about the identification with instanton partition function with surface operator insertions, we expect that for general affine level $\widehat{SL(3)}_k$ which corresponds to general values of $\epsilon_{1,2}$, the matching remains to work after restoring $\mathcal{K}$ factor into affine $SL(3)$ correlation function in \cite{Ribault:2008si}. Unfortunately apart from few known cases \cite{Alday:2010vg}, computing a closed expression for the effect of inserting $\mathcal{K}$ in affine conformal block is mathematically difficult. 
instead we generalized the result in \cite{Ribault:2008si} further by conjecturing that in \emph{critical/NS limit}, there is an exact matching between marked Toda conformal block, and affine conformal blocks \emph{without} $\mathcal{K}$ insertion, for arbitrary rank $N$ and higher genus. 
Here we explicitly demonstrate for the rank two case on genus one Riemann surface. While we have not been able to show explicitly, this matching combining with the gauge theory connection implies that the effect of ${\mathcal K}$ factor insertion on affine conformal block becomes negligible in the critical/NS limit. It would be very nice to verify this statement.

To demonstrate the matching of affine and Toda conformal block, we consider KZ and BPZ equations. For $A_{N-1}$ Toda theory the BPZ equation is of order $N$, thus one needs $N$th order generalization of KZ equation. As explained in \cite{Bernard198877}, the KZ equation arises as ``mixed Kac Moody-Virasoro'' Ward identity, where energy momentum tensor is identified with normal-ordered double currents operator. One can in principle obtain ``mixed Kac Moody-$\mathcal{W}_N$ algebra'' Ward identity if there is an $N$-current invariant in the Kac Moody algebra, and this will give rise to differential equation on conformal block of order $N$. 
For example, in \cite{Ribault:2008si} it is shown that due to cubic invariant in $SL(3)$ a third order differential equation is placed on affine $SL(3)$ correlation function. We expect those higher order differential operators to play the role of higher order conserved charges in integrable system. 

On the other hand, those higher order operators will match to the BPZ equations for higher rank Toda theory. Only for the rank one Liouville theory can one obtain a second order conserved charge, namely the Hamiltonian, from the BPZ equation such as in \cite{TakiMaruyoshi10}. To discuss integrable system, we therefore start from the original KZ equation, which always gives us second order Hamiltonian regardless of the rank.

\begin{figure}[h!]
\centering
\includegraphics[width=0.8\textwidth]{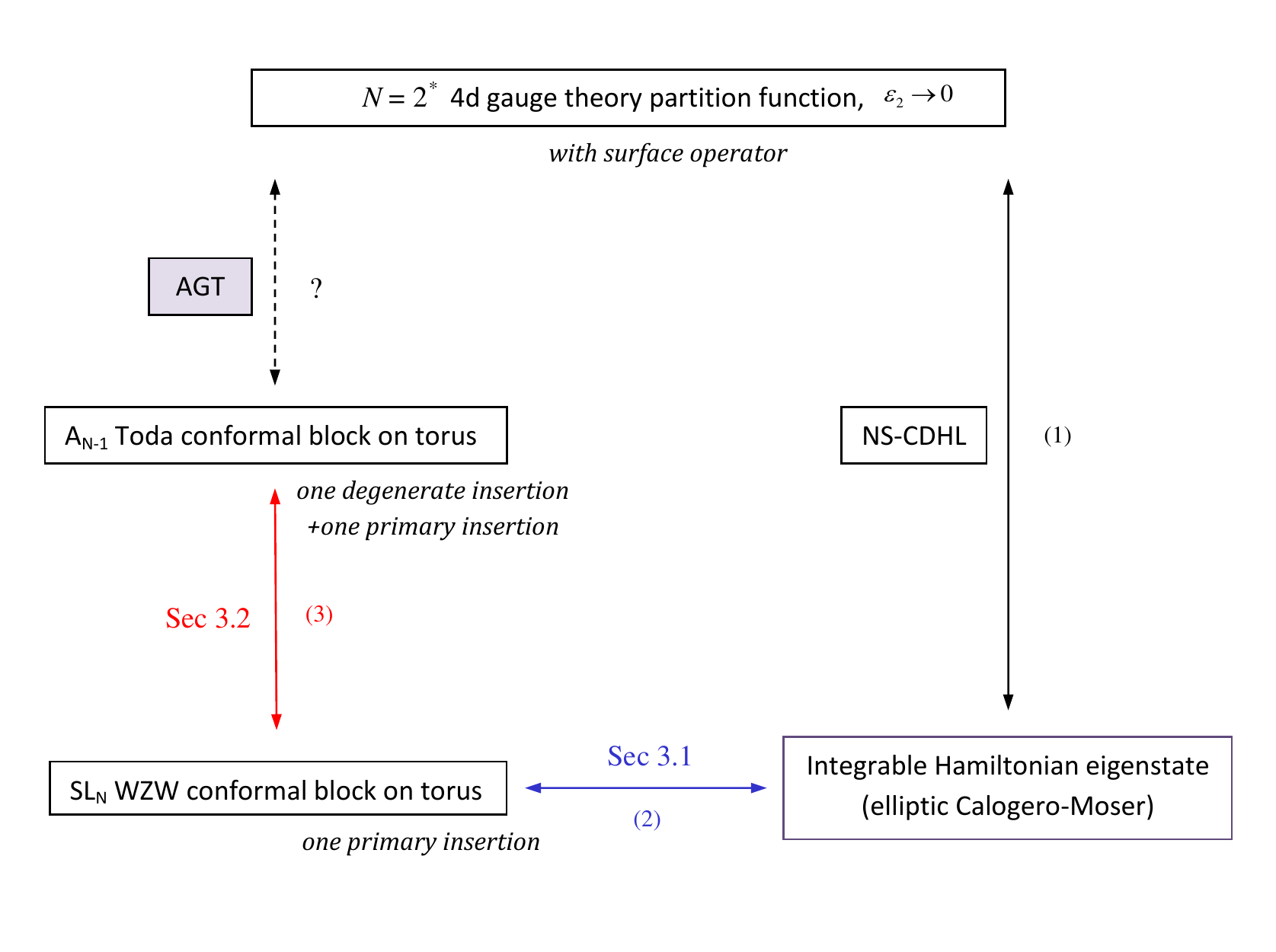}    
\caption{The chain of dualities related to four dimensional $\ssN=2^*$ theories.}
\label{fig:dualitytori}
\end{figure}

The logical flow of this section is summarized in \figref{fig:dualitytori}. In \secref{selfkzb}, we will demonstrate that the elliptic KZB equation constraining affine WZNW conformal block on the single punctured torus reduces to the Hamiltonian equation for elliptic Calogero system, thereby proving the connection between affine WZNW conformal block and elliptic Calogero-Moser eigenstate (arrow 2) \cite{Gorsky:1993dq}. On the other hand, it was shown in \cite{Nekrasov:2009rc,Chen:2011sj} that the four dimensional gauge theory partition functions in the NS limit can be identified with the eigenstates of the integrable systems, in particular for the five dimensional generalization of ${\mathcal N}=2^*$ theory, the corresponding integrable system was identified to be the elliptic Calogero-Moser system \cite{Chen:2012we}.

In \secref{DStorus}, we will establish the connection between affine $\mathfrak{sl}_3$ WZNW and $A_{2}$ Toda conformal blocks on the torus with appropriate degenerated insertions (arrow 3). This is achieved in the NS limit by matching the three-current KZB equations and third order BPZ equations constraining the corresponding conformal blocks. This equivalence between Toda and WZNW theory is also a manifestation of the Drinfeld-Sokolov reduction.

As a result, by connecting the chains of dualities (1)--(3) shown in \figref{fig:dualitytori} we can verify the AGT correspondence in NS limit.

\subsection{KZB Equation and Elliptic Calogero-Moser system}\label{selfkzb}
The generalization of KZ equation for affine WZNW theory on higher genus Riemann surface was made by Bernard \cite{Bernard198877}. It was later shown by Reshetikhin that the Knizhnik-Zamolodchikov-Bernard (KZB) system provides a quantization of the isomonodromy deformations equation \cite{KZReshet1992}. It was originally observed by Eguchi and Ooguri that the naive generalization on genus $g>0$ would contain expectation value of zero current mode and the equation is not self-contained \cite{Eguchi:1986sb}. The problem arises from the Ward identity of the current algebra on Riemann surface $\Sigma$ with genus $g>0$, schematically:
\begin{equation}\label{eqn:kzb0}
\langle J(z) \phi(w) \rangle = \mathcal{D} \langle \phi(w) \rangle + \sum_{j=1}^g \omega_j(z) \oint_{a_j} \langle J(u) \phi(w) \rangle \mathrm{d}u,
\end{equation}
where $\mathcal{D}$ is a differential operator, $J$ is the affine current, $\phi$ is a primary field, $\omega_i$ are set of 1-forms normalized under canonical basis in $H_1(\Sigma)$ as $\oint_{a_i}\omega_j=\delta_{ij}$, $\oint_{b_i}\omega_j=\tau_{ij}$, $\tau_{ij}$ is the period matrix. 

To obtain the constraint on correlation function one needs to know how current acts on correlation function, which can be obtained directly from equation (\ref{eqn:kzb0}) for the sphere case where the last term is absent, while for higher genus $g\geq 1$ with the last term the equation (\ref{eqn:kzb0}) cannot give us a direct constraint.

The problem was solved by Bernard \cite{Bernard198877} using insertion of extra group element in the conformal block and pull out the additional current in the last term by taking derivative in the algebra element. The ``twisting'' group element is labeled by the coordinates in the Cartan subalgebra,  specifically $\lambda=\sum_a\lambda_a \mathbf{h}_a\in \mathfrak{h}$, where $\mathbf{h}_a$ is an orthonormal basis of generators in Cartan sub-algebra. 

The KZB equation is most conveniently expressed in the form of \cite{Felder:1994gk}:
\begin{equation}\label{eqn:firstkzb}
\left[(k+h^{\vee})\frac{\partial}{\partial z_i}-\sum_{j\neq i}\Omega^{(i,j)}_z(z_i-z_j,\tau, \lambda)+\sum_a \mathbf{h}_a^{(i)} \frac{\partial}{\partial \lambda_a}\right]F(z,\lambda,\tau)=0,
\end{equation}
where we have also introduced the complex structure modulus $\tau$ of the torus and $\{z_i\}$ are the puncture positions.
This equation was obtained by varying conformal block with respect to the coordinate of puncture $z_j$ using $\widehat{\mathcal{L}}_{-1}=\partial_z$. 
$\Omega_z$ is the $z$ component for 1-form $\Omega$ with coordinate $(z,\tau)$, which has the following expression: 
\begin{align}\label{omegaform}
\Omega(z,\lambda,\tau)\equiv & \omega_0(z,\tau) C_0 + \sum_{\alpha}\omega_{1,\alpha(\lambda)}(z,\tau)C_{\alpha}\\
                         =   & \Omega_z \mathrm{d}z + \Omega_{\tau} \mathrm{d}\tau,
\end{align}
where $C_{\alpha}=e_{\alpha}e_{-\alpha}$, with $e_{\alpha}$ a basis for root subspace $\mathfrak{g}_\alpha=\{g\in\mathfrak{g}\;|\;[\mathbf{h},g]=\alpha(\mathbf{h})g\;\forall \mathbf{h}\in\mathfrak{h}\}$. $C_0 = \sum_a \mathbf{h}_a \mathbf{h}_a$ is the second symmetric invariant tensor of Cartan subalgebra.
The 1-form coefficients in (\ref{omegaform}) are:
\begin{align}
&\omega_0(z,\tau)= \rho(z,\tau)\mathrm{d}z+\frac{1}{2}\left(\rho(z,\tau)^2+\partial_z \rho(z,\tau)\right) \mathrm{d}\tau,\label{eqn:function1}\\
&\omega_{1,w}(z,\tau)= \sigma_{w}(z,\tau)\mathrm{d}z  - \partial_w \sigma_w(z,\tau) \mathrm{d}\tau,\label{eqn:function2}\\
&\rho(z,\tau)\equiv\frac{\vartheta_1'(z|\tau)}{\vartheta_1(z|\tau)},
\quad\vartheta_1(z|\tau)\equiv -\sum_{l\in \mathbb{Z}}e^{\pi i\tau\left(l+\frac{1}{2}\right)^2+2\pi i\left(l+\frac{1}{2}\right)\left(z+\frac{1}{2}\right)}\label{eqn:theta},\\
&\sigma_w(z,\tau)\equiv\frac{\vartheta_1(w-z|\tau)\vartheta_1'(0|\tau)}{\vartheta_1(w|\tau)\vartheta_1(z|\tau)}.
\end{align}
One can also vary the correlation function with respect to modular parameter $\tau$, which results in \cite{Felder:1994gk}:
\begin{equation}\label{eqn:kzbheat}
\left[\sum_a \frac{\partial^2}{\partial\lambda_a^2}+\sum_{i,j}\Omega^{(i,j)}_{\tau}(z_i-z_j,\tau,\lambda)-4\pi i (k+h^{\vee})\partial_\tau \right]F=0.
\end{equation}
Equation (\ref{eqn:kzbheat}) is conventionally called KZB heat equation.
The compatibility of equation (\ref{eqn:firstkzb}) and (\ref{eqn:kzbheat}) can be put into the commutation relation
\begin{equation}
[D_1,\;D_2 ]=0
\end{equation}
where $D_1 F =0$ and $D_2 F=0$ correspond to KZB equation (\ref{eqn:firstkzb}) and KZB heat equation (\ref{eqn:kzbheat}) respectively. This equation was expressed by Felder and Varchenko in \cite{IntegralrepKZB} as the flatness condition of KZB connection, and was explicitly proved.


Now consider affine conformal block with only one insertion.
The KZB heat equation takes the form of non-stationary Schr\"odinger equation:
\begin{equation}\label{eqn:KZB}
\left[ \sum_b \frac{\partial^2}{\partial \lambda_b^2}-\eta_1 C_0-\sum_{\alpha\in\Delta}\wp\left(\alpha(\lambda)|\tau\right) e_{\alpha} e_{-\alpha} \right]F(z, \lambda,\tau)=4\pi i (k+h^{\vee}) \partial_{\tau} F(z,\lambda,\tau).
\end{equation}
For $\mathfrak{g}=\mathfrak{sl}_N$ and primary operator in spin $j_e$ representation, we recover
\begin{align}\label{eqn:kzbecm}
&\left[-\frac{1}{2}\sum_a \frac{\partial^2}{\partial \lambda_a^2}+j_e(j_e+1)\sum_{a>b}\left( \wp(\lambda_a-\lambda_b|\tau)+2\eta_1(\tau) \right)\right] F(z,\lambda,\tau)\cr
&=-2\pi i (k+N) \partial_{\tau} F(z,\lambda,\tau)\,,
\end{align}
where $k=-N-b^{-2}$, $\wp(z|\tau)$ is the Weierstrass $\wp$-function, $\vartheta_1$ is defined in (\ref{eqn:theta}),
and $\eta_1(\tau)\equiv -\frac{1}{6}\frac{\vartheta_1'''(0|\tau)}{\vartheta_1'(0|\tau)}$.

Now we will take critical level limit of equation (\ref{eqn:kzbecm}). For $SL(N)$, we expect from \cite{Alday:2010vg} that torus one point conformal block to have the divergent structure in the critical level limit as following:
\begin{equation}\label{eqn:nsdiv}
F=q^{\frac{1}{k+N} j_i(j_i+1)}e^{\frac{1}{k+N} W(q)}F_{\text{reg}}(z_1,\lambda,\tau),
\end{equation}
where $q\equiv e^{2\pi i \tau}$, $W(q)$ is some function of $q$ finite in the critical limit, and $j_i$ labels the Hilbert space. The parameters can be matched to the gauge theory as
\begin{equation}
j_e=-N \frac{m}{2\epsilon_1},\quad j_i=-\frac{1}{2}+\frac{a_i-a_{i+1}}{2\epsilon_1},\quad k=-N-\frac{\epsilon_2}{\epsilon_1},\quad \tau=\frac{\theta}{2\pi}+\frac{4\pi i}{g^2},
\end{equation}
where $m$ is the adjoint mass, $a_i$ are Coulomb parameters, and $\epsilon_{1,2}$ are equivariant parameters; the torus complex structure $\tau$ is matched to the gauge theory complexified coupling $\frac{\theta}{2\pi}+\frac{4\pi i}{g^2}$.
Upon substitution, the KZB heat equation for $SL(N)$ on regularized conformal block has the form
\begin{align}
&\left[-\frac{1}{2}\sum_a \frac{\partial^2}{\partial \lambda_a^2}+j_e(j_e+1)\sum_{a>b}\left( \wp(\lambda_a-\lambda_b|\tau)+2\eta_1(\tau) \right)\right] F_{\text{reg}}(z_1,\lambda,\tau)\nonumber\\
&\qquad=4\pi^2 \left[  j_i(j_i+1)+q W'(q) \right]F_{\text{reg}}(z_1,\lambda,\tau) - 2\pi i (k+N)\partial_{\tau}F_{\text{reg}}(z_1,\lambda,\tau).
\end{align} 
In the NS limit $\epsilon_2\to 0$, it becomes
\begin{align}
&\left[-\frac{1}{2}\sum_a \frac{\partial^2}{\partial \lambda_a^2}+j_e(j_e+1)\sum_{a>b}\left( \wp(\lambda_a-\lambda_b|\tau)+2\eta_1(\tau) \right)\right] F_{\text{reg}}(z_1,\lambda,\tau)\nonumber\\
& \qquad =4\pi^2 \left[  j_i(j_i+1)+q W'(q) \right]F_{\text{reg}}(z_1,\lambda,\tau).\label{eqn:ecmkzbb}
\end{align}
Thus at critical level limit one point conformal block become eigenstate of elliptic Calogero-Moser system.

\bigskip

One may wonder what is the role of the first KZB equation (\ref{eqn:firstkzb}). In fact, below we will show that equation (\ref{eqn:firstkzb}) reduces trivially in the case of one puncture.

First note that the position of one puncture on torus depends only on its complex structure $\tau$ therefore the first term in (\ref{eqn:firstkzb}) drops off. The second term in equation (\ref{eqn:firstkzb}) also vanishes. To see this, note the argument of $\Omega$ in the KZB equations depends on difference of puncture positions, and for one puncture case it amounts to setting argument $z=0$, and $z$ component of $\Omega$ vanishes:
\begin{equation}\nonumber
\Omega_z(0,\lambda,\tau)= \rho(z,\tau)|_{z\to 0} \left( C_0 + \sum_{\alpha}C_{\alpha}\right) = 0 \cdot \left( C_0 + \sum_{\alpha}C_{\alpha}\right) =0,
\end{equation}
leaving only $\tau$ component $\Omega_{\tau}(0,\lambda,\tau) = -\left(\eta_1C_0+\sum_{\alpha}\wp(\alpha(\lambda))e_{\alpha}e_{-\alpha} \right)$. 
Note that $\rho(z,\tau)\equiv\frac{\vartheta_1'(z|\tau)}{\vartheta_1(z|\tau)}$ vanishes at $z \to 0$ since the Jacobi theta function $\vartheta_1(z|\tau)$ is odd in $z$.

To kill the last term in (\ref{eqn:firstkzb}), we need the neutrality condition in \cite{Bernard198877}.
Consider twisted correlation function inserted with Cartan subgroup element $g$: $F = \langle \prod_{i}\phi_i(z_i)\rangle_g$, which is invariant under the conjugation of Cartan group element $g\mapsto g' g g'^{-1}=g$ since elements in Cartan subgroup commute with each other. The invariant property can be expressed as: \cite{Bernard198877}
\begin{align}\label{eqn:neutral}
\left\langle \prod_{i}\phi_i(z_i)\right\rangle_{g'gg'^{-1}}=&\left(\prod_i\rho_i(g')\right)\left\langle\prod_j \phi_j(z_j)\right\rangle_{g}=\left\langle \prod_{j}\phi_j(z_j)\right\rangle_{g},
\end{align}
where both $g$ and $g'$ belong to maximal torus $T$, or Cartan subgroup. $\rho_i$ is the representation of $T$ by field $\phi_j$, $g\cdot\phi_j(z_j)\cdot g^{-1}=\rho_j(g)\cdot \phi_j(z_j)$. 

In particular, take $g'$ to be one of the Cartan generator such that $g'=\exp(\epsilon \mathbf{h}_a)$, in the infinitesimal form the equation (\ref{eqn:neutral}) becomes
\begin{equation}\label{eqn:charge0}
\left(\sum_i \mathbf{h}_a^{(i)}\right) \left\langle\prod_j \phi_j(z_j)\right\rangle_{g}=0
\end{equation}
for each generator $\mathbf{h}_a$ of Cartan subalgebra and all elements $g$ in maximal torus $T$. We will call equation (\ref{eqn:charge0}) the ``neutral condition''.

In the case of one puncture conformal block, the conjugation invariant property of twisted conformal block, expressed as neutral condition in equation (\ref{eqn:charge0}), becomes: (rewrite $g=\lambda$) 
\begin{equation}
\mathbf{h}_a^{(1)} \langle \phi_1(z_1)\rangle_{\lambda}=0=\mathbf{h}_a^{(1)}F
\end{equation}
for all Cartan generators $h_a$, therefore the last term in the first KZB equation (\ref{eqn:firstkzb}) drops off, and (\ref{eqn:firstkzb}) is completely killed in the one puncture case.

In summary, we have shown the KZB equations, which constrain affine WZNW conformal block, reduced to one elliptic Calogero-Moser eigen equation (\ref{eqn:ecmkzbb}) in the NS limit. Therefore the conformal block and eigenstate match, giving arrow (2) in \figref{fig:dualitytori}.

\subsection{Matching of WZNW and Toda Conformal Blocks on the Torus}\label{DStorus}
%
In this section we will show the arrow (3) in Figure \ref{fig:dualitytori}, namely the connection between $A_{N-1}$ Toda and affine $SL(N)$ WZNW theories on torus. 
For definiteness, we will show the rank 2 case for $SL(3)$ and $A_2$ Toda theory. We will derive the analogue of BPZ equation on torus satisfied by $A_2$ Toda conformal block, and match the BPZ equation with KZB equation, thereby verifying the following equality for $A_2$ Toda and SL(3) WZNW conformal blocks at critical level limit $k \to -3$: \cite{Ribault:2008si, H3+1}:
\begin{equation}
\langle V_j(z)\rangle_{(\lambda,\tau)}=\Theta(y,z,\tau) \langle V_{3,1}(y)V_{\alpha}(z)\rangle,
\label{eqn:matching}
\end{equation}
where $\Theta(y,z,\tau)$ is the twisting factor, and $\lambda$ is the twisting parameter. In the following we will take $z=0$ without lost of generality. The angle bracket is understood to be normalized conformal block.

\subsubsection{$\mathcal{W}_3$ BPZ equation for degenerate Toda field on torus}
The $A_2$ Toda theory enjoys $\mathcal{W}_3$ symmetry in addition to Virasoro algebra, which has the following commutator relations \cite{FateevLitvinovToda}

\begin{align}
[W_m,W_n]=& (m-n)\left(\frac{1}{15}(m+n+2)(m+n+3)-\frac{1}{6}(m+2)(n+2)\right)L_{m+n}+\nonumber\\
          & \quad \frac{c}{3\cdot 5!}(m^2-1)(m^2-4)m\delta_{m+n,0}+\frac{16}{22+5c}(m-n)\Lambda_{m+n},\label{eqn:wcomm}\\
[L_m,W_n]=&(2m-n)W_{m+n},\\
[L_m,L_n]=& (m-n)L_{m+n}+\frac{c}{12}(m^3-m)\delta_{m+n,0},
\end{align}
where we denote
\begin{equation}
\Lambda_n=\sum_{j\in \mathbb{Z}}:L_j L_{n-j}: +\frac{1}{5}x_n L_n \text{ with } \left\{ \begin{array}{l}
x_{2l}=(1+l)(1-l)\\
x_{2l+1}=(2+l)(1-l) \end{array}
\right. ,
\end{equation}
and the normal ordering $:L_{n}L_{m}:=L_{<}L_{>}$, where $>$ denotes the largest number among $m$,$n$, etc.
The central charge for $\mathcal{W}_3$ is given in \cite{FateevLitvinovToda} as $c=2\left(1+12(b+b^{-1})^2\right)$.

Define the primary field $\phi$ as following:
\begin{align*}
&L_n \phi=0\text{,\, } W_n\phi=0\quad \text{for} \quad n>0;\\
&L_0\phi=h\phi\text{,\, }W_0\phi=v\phi,
\end{align*}
where $h$ and $v$ are the weight of this primary field with respect to Virasoro and $Wn$ algebra.


It is known that $\mathcal{W}_3$ algebra admits a third level null state with the following weight data:\cite{FateevLitvinovToda}
\begin{subequations}
\begin{equation}\label{deg1}
    h=-\frac{4b^2}{3}-1\qquad
    v^2=-\frac{2h^2}{27}\frac{5b+\frac{3}{b}}{3b+\frac{5}{b}},
\end{equation}
\begin{equation}\label{deg2}
    h=-\frac{4}{3b^2}-1\qquad
    v^2=-\frac{2h^2}{27}\frac{3b+\frac{5}{b}}{5b+\frac{3}{b}}.
\end{equation}
\end{subequations}
The two equations here are related through $b \to 1/b$ transformation and they give rise to a set of third order null equations on degenerate field $V_{3,1}$\cite{FateevLitvinovToda}
\begin{align}
\left(W_{-1}-\frac{3v}{2h} L_{-1} \right)V_{3,1} & = 0,\label{eqn:null1}\\
\left(W_{-2}-\frac{12v}{h(5h+1)} L_{-1}^2+
    \frac{6v(h+1)}{h(5h+1)}L_{-2 }\right)V_{3,1} & = 0, \label{eqn:null2}\\
\left(W_{-3}-\frac{16v}{h(h+1)(5h+1)}
   L_{-1}^3+\frac{12v}{h(5h+1)} L_{-1} L_{-2}+
    \frac{3v(h-3)}{2h(5h+1)} L_{-3}\right)V_{3,1} & = 0.\label{eqn:null3}
\end{align}
Taking the commutator of the equations (\ref{eqn:null1}) and (\ref{eqn:null2}),
one finds
\begin{align}\nonumber
\left\{\frac{32}{22+5c}\left(L_{-3}L_0+L_{-2}L_{-1}\right)-\frac{24v}{h(5h+1)}L_{-1}W_{-2}+\frac{18v(h+1)}{h(5h+1)}W_{-3}-\frac{9v^2(h+1)}{h^2(5h+1)}L_{-3}
\right\}V_{3,1}&\\ \notag
=0\,.&
\end{align}
After substituting the $\mathcal{W}_3$ generators using equations (\ref{eqn:null1})-(\ref{eqn:null3}), the algebraic null equation becomes
\begin{equation}
\left\{c_1 L_{-1}L_{-2}+c_2 L_{-3}\right\}V_{3,1}=0
\label{eqn:W3nullnull}
\end{equation}
with
\begin{align}
&c_1=\frac{4b^2}{15+34b^2+15b^4}-\frac{72v^2(h+1)}{h^2(5h+1)^2},\nonumber\\
&c_2=\frac{4b^2(h-1)}{15+34b^2+15b^4}-\frac{72v^2(h^2-1)}{h^2(5h+1)^2}.\nonumber
\end{align}
One would thus expect the BPZ equation obtained to be a second order differential equation as for Liouville field theory. However, after substituting degenerate weights (\ref{deg1}) or (\ref{deg2}), equation (\ref{eqn:W3nullnull}) vanishes identically, therefore one has to seek a higher order BPZ equation. 

One can obtain a third order BPZ equation by inserting the equations (\ref{eqn:null1})-(\ref{eqn:null3}) into correlation functions with the aids of Virasoro and $\mathcal{W}_3$ Ward identity. Virasoro Ward identity on torus is available for example in \cite{Eguchi:1986sb}, while $\mathcal{W}_3$ Ward identity is lacking in the literature, below we will propose a $\mathcal{W}_3$ Ward identity from OPE and modular invariant.

Note the $\mathcal{W}_3$ generator and primary $V$ has the following OPE structure:
\begin{equation}
W(y)V(z)=\frac{W_{0}V(z)}{(y-z)^3}+\frac{W_{-1}V(z)}{(y-z)^2}+\frac{W_{-2}V(z)}{y-z}+\text{reg}.
\end{equation}
When taking expectation value on torus the regular terms also contribute in addition to singular terms in OPE. To have the correct pole structure and keep modular invariant, a natural guess for the $\mathcal{W}_3$ Ward identity would be
\begin{align}
&\langle W(y)V_1(z_1)\cdots V_n(z_n)\rangle \nonumber \\
&= \left\{ \sum_{i=1}^n \left[-\frac{1}{2}\wp'(y-z_i|\tau)\hat{W}_{0}V(z_i) + \wp(y-z_i|\tau) \hat{W}_{-1}V(z) + 
\zeta(y-z_i|\tau)\hat{W}_{-2}V(z) \right] +\mathcal{O}_{\tau}\right\}\nonumber\\
&\qquad\qquad\qquad\qquad\qquad\qquad\qquad\qquad\qquad\qquad\qquad\qquad\qquad\qquad\qquad \langle V_1(z_1)\cdots V_n (z_n)\rangle\label{eqn:W3Ward}
\end{align}
where $\zeta$ and $\wp$ are elliptic functions in \cite{Eguchi:1986sb} with the pole structures $\zeta(z)\sim 1/z +c_1(\tau)z^3+\cdots$ and $\wp(z)\sim 1/z^2+c_2(\tau)z^2+\cdots $. 
$\mathcal{O}_{\tau}$ is an operator containing only $\tau$.

One can then insert the null equations (\ref{eqn:null1})-(\ref{eqn:null3}) into Toda two-point toric correlation function using
\begin{align*}
&L_{-s} V(z) = \frac{1}{2\pi i}\oint_{\mathcal{C}(z)}\frac{\text{d}w}{(w-z)^{s-1}}T(w)V(z)\\
&W_{-s} V(z) = \frac{1}{2\pi i}\oint_{\mathcal{C}(z)}\frac{\text{d}w}{(w-z)^{s-2}}W(w)V(z),
\end{align*}
they give rise to BPZ equation:
\begin{align}
&\langle \hat{W}_{-3} V_{3,1}(y)V(0)\rangle\nonumber\\
&\qquad = \left[ -\frac{1}{2}\wp'(y|\tau) \hat{W}_{0}^{(2)} + \wp(y|\tau)\hat{W}_{-1}^{(2)}+ 
\zeta(y|\tau) \hat{W}_{-2}^{(2)}+ \mathcal{O}_{\tau} \right] \langle V_{3,1}(y) V(0)\rangle  \nonumber\\ 
&\qquad = 3v\left\{ -b^2 \frac{\partial^3}{\partial y^3}-\frac{\partial}{\partial y}\left[ 2\pi i K_{\tau}-b^2 C_2(j)\left(\wp(y|\tau)+2\eta_1\right)+\left(-\zeta(y|\tau)+2\eta_1 y\right)\frac{\partial}{\partial y}\right]\right.+\nonumber\\
&\qquad\quad\left. \quad \frac{1}{2}\left[-b^2C_2\wp'(y|\tau)+\wp(y|\tau)\frac{\partial}{\partial y}\right]\right\} \langle V_{3,1}(y)V(0)\rangle\,,\nonumber
\end{align}
which is essentially equation (\ref{eqn:null3}). The superscript $\hat{W}_{0}^{(2)}$ denotes acting on second insertion $V(0)$. Since we are considering NS limit $b\to \infty$ we have dropped some of lower order terms in $b$.
Hereafter we will only retain leading terms of order $\mathcal{O}(b^2)$.  
The above equation can be further simplified to
\begin{align}\label{eqn:3JBPZmatch}
&\langle \hat{W}_{-3} V_{3,1}(y)V(0)\rangle\nonumber\\
&\qquad = \left[ -\frac{1}{2}\wp'(y|\tau) \hat{W}_{0}^{(2)} + \wp(y|\tau)\hat{W}_{-1}^{(2)}+ 
\zeta(y|\tau) \hat{W}_{-2}^{(2)}+ \mathcal{O}_{\tau} \right] \langle V_{3,1}(y) V(0)\rangle  \nonumber\\ 
&\qquad = i\sqrt{\frac{2}{5}}\left\{ b^2\left(\frac{1}{2}\frac{\partial^3}{\partial y^3}+\frac{1}{2}\frac{\partial}{\partial y}
\left[\frac{\partial^2}{\partial y^2}-C_2\left(\wp(y|\tau)+2\eta_1\right)+4\pi i b^{-2} K_{\tau}\right] \right)+\mathcal{O}(1)\right\} \\\nonumber
&\qquad \langle V_{3,1}(y)V(0)\rangle\,.		
\end{align}
Observe that at the third line the operator in the square brackets are just second order KZB operator and will vanish when acting on the corresponding affine WZNW toric conformal block. Also note that the second line is just $\mathcal{W}_3$ Ward identity operator acting on $\langle V(0)\rangle_{\text{affine}}$.


\subsubsection{Three-current Affine KZB Equation}
Since the BPZ equation \eqref{eqn:3JBPZmatch} we obtained is of third order, we have to find a three-current generalization of the KZB equation in section (4.1). This is achieved by inserting the following relation between $\mathcal{W}_3$ generator and affine Casimir operator \cite{Bouwknegt:1992wg}:
\begin{equation}
W(y)=\frac{1}{6}\eta^{(3)}(k)d^{abc} \left(J^a \left( J^b J^c\right)\right)(y),
\label{eqn:W3JJJ}
\end{equation}
where $d^{abc}=\text{tr}(t^at^bt^c+t^at^ct^b)$, and
$\eta^{(3)}(k)=\frac{1}{k+3}\sqrt{\frac{6}{5(2k+3)}} \sim i b^2 \sqrt{\frac{2}{5}}$ in the critical level limit $k\to -3$.

Insert equation \eqref{eqn:W3JJJ} into affine conformal block $\langle V(z) \rangle_{\text{affine}}$ we have 
\begin{align}
&\langle W(y) V(z)\rangle\nonumber\\
&\qquad = \left[ -\frac{1}{2}\wp'(y-z|\tau) \hat{W}_{0} + \wp(y-z|\tau)\hat{W}_{-1} + 
\zeta(y-z|\tau)\hat{W}_{-2}+ \mathcal{O}_{\tau}\right] \langle V(z)\rangle \nonumber\\ 
&\qquad = i b^2 \sqrt{\frac{2}{5}} \frac{d_{abc}}{6} \langle \left(J^a \left( J^b J^c\right)\right)(y)V(z)\rangle + \mathcal{O}(1).
\end{align}
Hereafter we will set the vertex insertion position to be $z=0$ without lost of generality.

The last line in the previous equation can be obtained using the method in section 4 of \cite{Eguchi:1986sb}:
\begin{align}
&\left\langle \left(J^a \left( J^b J^c\right)\right)(y)V(0)\right\rangle=\nonumber\\
&\prod_{l=a,b,c} \left[\left(\zeta(y|\tau)-2\eta_1 y\right)t^l+\frac{\partial}{\partial \lambda_l}\right]\nonumber\\
&\qquad-2\eta_1\left( K^{ab}\left[\left(\zeta(y|\tau)-2\eta_1y\right)t^c+  \frac{\partial}{\partial \lambda_c}\right]+\text{cycl.}\right),\nonumber
\end{align}
where $K^{ab}=\text{tr}(t^at^b)\propto \delta^{ab}$. (Note that $K^{ab}d^{abc}\propto \text{tr}(t^at^at^c)= \text{tr}(C_2t^c)=0$.)

Therefore we have the three-current KZB equation
\begin{align}
&\left[ -\frac{1}{2}\wp'(y|\tau) \hat{W}_{0} + \wp(y|\tau)\hat{W}_{-1} + 
\zeta(y|\tau)\hat{W}_{-2}+ \mathcal{O}_{\tau}\right] \langle V(0)\rangle \nonumber\\ 
&\qquad = i b^2 \sqrt{\frac{2}{5}} \frac{d_{abc}}{6} \prod_{l=a,b,c} \left[\left(\zeta(y|\tau)-2\eta_1 y\right)t^l+\frac{\partial}{\partial \lambda_l}\right]\langle V(0)\rangle.\label{eqn:3JKZBmatch}
\end{align}

\subsubsection{Twisting factor}
The twisting factor appears in equation (\ref{eqn:matching}) can be obtained from free field correlation function \cite{Ribault:2008si}\cite{H3+1}, which has the following ansatz:
\begin{equation}
\Theta\sim \prod_{i<j} f(y_i-y_j)^{\alpha}\prod_{k<l} f(z_k-z_l)^{\beta} \prod_{i,k}f(y_i-z_k)^{\gamma},
\end{equation}
where the function $f$ depends on genus (e.g. id for sphere and Jacobi theta function for torus), but the constants $\alpha, \beta,\gamma$ come from matching of weights and OPE pole structure etc. and depends only on the algebra, therefore the same for any genus.
The twisting factor for SL(3) on torus can thus be deduced using $f$ on torus in \cite{H3+1} and $\alpha,\beta,\gamma$ for SL(3) on sphere in \cite{Ribault:2008si}, as follows:
\begin{equation}\label{eqn:twistfact}
\Theta=C\theta(y-z|\tau)^{\frac{1}{b^2}}\,.
\end{equation}
Note that we have a sign difference to \cite{Ribault:2008si} in the exponent since we use a different convention between $b^2$ and level $k$ following \cite{Alday:2010vg}.

\subsubsection{Separation of Variables}
We observe that both the BPZ equation and KZB equation satisfied by separate conformal blocks containing Weierstrass elliptic function $\wp$. In particular the three-current BPZ equation (\ref{eqn:3JBPZmatch}) contains a second KZB operator with potential $\wp(y|\tau)+2\eta$, which is to be matched with the KZB operator we obtained in section (4.1), with potential $\wp(\lambda_2-\lambda_1|\tau)+2\eta$. Therefore in order to match the conformal blocks, it is natural to identify $y=\lambda_2-\lambda_1=\vec{\lambda}\cdot (\mathbf{h}_2-\mathbf{h}_1)$, where $\mathbf{h}_a$ is some basis in Cartan subalgebra.

In addition, in \cite{H3+1} the vector $\frac{\partial}{\partial y}$ is matched to $\frac{\partial}{\partial \lambda}-J^0$, which when acting singly annihilates affine conformal block. In our case we therefore propose that when acting on affine conformal block the vector $\frac{\partial}{\partial y}$ should match to $\mathbf{h}_a\frac{\partial}{\partial \lambda^a}$, which also annihilates affine conformal block due to neutrality condition. In particular the triple product in equation (\ref{eqn:3JKZBmatch}) is to be matched to $\frac{\partial^3}{\partial y^3}$ following \cite{H3+1}.

\subsubsection{Matching}
In this subsection we will verify the relation (\ref{eqn:matching}). Note that three-current KZB equation tells us $D^{(\text{3jKZB})}\langle V_j(0)\rangle_{(\lambda,\tau)}=0$, on the Toda CFT side after including twisting factor this becomes:
\begin{equation}\nonumber
D^{(\text{3jKZB})}\Theta(y,0,\tau) \langle V_{3,1}(y)V_{\alpha}(0)\rangle =0,
\end{equation}
where the operators can be combined to give a third order differential operator that annihilate Toda conformal block, but this by definition should just be the third order BPZ operator from equation (\ref{eqn:3JBPZmatch}), denoted as $H^{(3)}$. Therefore, our aim is to match $\left( D^{(\text{3jKZB})}\Theta(y,0,\tau) \right)$ with $H^{(3)}$, or turning the logics other way around, matching 
\begin{equation}
H^{(3)}\Theta^{-1}
\end{equation}
with three-current KZB operator (\ref{eqn:3JKZBmatch}) in the NS limit, up to some overall constant. Note that at the leading order $\mathcal{O}(b^2)$ the effect of twisting operator can be neglected. Using the matching prescription in previous section, the three-current KZB operator becomes
\begin{align}
&\left[ -\frac{1}{2}\wp'(y|\tau) \hat{W}_{0} + \wp(y|\tau)\hat{W}_{-1} 
\zeta(y|\tau)\hat{W}_{-2}+ \mathcal{O}_{\tau}\right]- i \frac{1}{2} b^2 \sqrt{\frac{2}{5}} \frac{\partial^3}{\partial y^3},
\end{align}
which coincides at $\mathcal{O}(b^2)$ with the third order BPZ operator
\begin{align}
&\left[ -\frac{1}{2}\wp'(y|\tau) \hat{W}_{0}+ \wp(y|\tau)\hat{W}_{-1}+ 
\zeta(y|\tau) \hat{W}_{-2}+ \mathcal{O}_{\tau} \right]\nonumber\\ 
&-i\frac{1}{2}  b^2\sqrt{\frac{2}{5}}   
	\left\{
	\frac{\partial^3}{\partial y^3}+
					\frac{\partial}{\partial y}
					\left[\frac{\partial^2}{\partial y^2}-C_2\left(\wp(y|\tau)+2\eta_1\right)+4\pi i b^{-2} K_{\tau}\right] \right\}.
\end{align}
up to the two-current KZB operator that annihilate the conformal block. Therefore, the WZNW and Toda conformal blocks match at NS limit.

\subsection{Generalization to ADE Gauge Groups}
Before ending this section we like to comment that the proof of AT duality we presented in section (\ref{selfkzb}) can be generalized straightforward to ADE type gauge group and affine CFT on torus. First note that the one-puncture KZB heat equation (\ref{eqn:KZB}) is valid for any complex simple Lie algebra $\mathfrak{g}$. If we assume the affine-$\mathfrak{g}$ conformal block has the same divergence structure near critical level as in equation (\ref{eqn:nsdiv}) with $j_i(j_i+1)$ replaced by second Casimir, then from the same argument before the affine one-puncture elliptic conformal block should coincide with eigenstate of quantum elliptic $\mathfrak{g}$-Calogero system:
\begin{equation}\label{eqn:eCMg}
\widehat{H}^{eCM}= -\frac{1}{2}\frac{\partial^2}{\partial\lambda^2}+\sum_{\alpha\in\Delta(\mathfrak{g})}\frac{1}{2}m_{\alpha}^2\wp(\alpha(\lambda)|\tau),
\end{equation}
where $m_{\alpha}^2 = e_{\alpha}e_{-\alpha}$ depends only on the Weyl orbit.

On the other hand D'Hoker and Phong have shown in \cite{D'Hoker:1998yi} that integrable system of equation (\ref{eqn:eCMg}) with simply-laced (i.e. ADE type) algebra $\mathfrak{g}$ can give rise to Seiberg-Witten curve and associated differential for $\mathcal{N}=2^*$ four dimensional supersymmetric gauge theory. Therefore we expect from section (\ref{selfkzb}) that ADE-type $\mathcal{N}=2^*$ gauge theory partition function with a surface operator will obey affine-ADE algebra, which is a generalized AT correspondence on torus for all single ADE type gauge group.

Given this generalized AT duality, and assuming our result for CFTs resulting from surface operator also holds for the generalized algebra, we can establish a proof for AGT relation on torus with ADE type gauge group.

\section{Future Directions}\label{sec:Future}
In this paper we used the equivalence between two different quantizations of the vacua moduli space of conformal 4d theories in order to prove the bispectral duality between the XXX and trigonometric Gaudin models. It would be interesting to develop a more systematic approach to the duality, similar to the one in \cite{Gaiotto:2013bwa}, but adopted to the noncompact symmetries. In particular the 5d/3d duality considered in \cite{Chen:2012we} may be of the great help. Indeed, the XXZ $SL(N)$ chains will help their S-dual counterparts, and, supposedly, XXX/tGaudin pair will follow by taking the proper limit. It is certainly worthwhile extending the analysis we have carried out in this work to other quiver gauge theories in 4d which have string theory origin \cite{Orlando:2011nc}. 

We also found that the two quantizations coincide for the genus one spectral curve and thus identifies quantum elliptic Calogero-Moser systems. It would be interesting to investigate these phenomena further for higher genera and their implication to integrable systems. The different quatizations can also potentially be related to the different descriptions of surface operator in gauge theory from 6d $\mathcal{N}=(0,2)$ theory. The latter is illustrated in the paper by matching the associated Toda and affine conformal blocks at the NS/critical level limit. It may be interesting to gain a further understanding on the role of surface operator plays in this Drinfeld reduction-like scenario and duality between quantum integrable systems by going beyond the NS limit.

As far as generic D and E type quivers (as well as their affine versions) are concerned, there is no brane description of the Coulomb/Higgs branes of the theories available. It certainly does not imply that the 2d GLSM cannot be derived, only that it might be quite 
cumbersome to understand what the corresponding vortex moduli space is. However, the quantum version of the Nekrasov-Pestun solution will soon be available for all ADE quivers \cite{NP2}, and the missing quantum models will be identified. Presumably, those new models will be bispectrally dual to the quantized integrable systems emanating from generic D and E quiver theories in 4d \cite{Nekrasov:2012xe}. The details of this anticipated duality are very intriguing. 

In this work we studied only a restricted class of the Gaudin systems, namely when only two Lax matrices had the maximal rank, and the others had unit rank. This choice was dictated by the Lagrangian nature of the UV 4d quiver theories we have started with. Indeed, in order to get a 4d theory which admits a Lagrangian description, one wraps M5 branes around the Riemann surface with such punctures. More generic monodromies at punctures would lead to a genuinely non-Lagrangian theory.
Yet, the Hitchin system on a sphere with generic monodormies makes perfect sense. It would be interesting to study its potential bispectral dual. This duality should be related to some interesting symmetry of the 6d $(0,2)$ theory which is yet to be discovered.

%
%

\section*{Acknowledgments}
We are grateful to Davide Gaiotto, Nikita Nekrsov, Vasya Pestun, Andrey Zotov for very helpful and interesting discussions, and especially to Sasha Gorsky, who collaborated with us on the initial stage of the project.
The research of HYC is supported in part by National Science Council through the grant No.101-2112-M-002-027-MY3 and Center for Theoretical Sciences at National Taiwan University. PK is thankful to Simons Center for Geometry and Physics at Stony Brook University as well as to W. Fine Institute for Theoretical Physics at University of Minnesota, where part of his work was done, for kind hospitality. HYC would also like to thank the generous support from Kenda Foundation. Research of PK at University of Minnesota was supported in part by the National Science Foundation under Grant No.\! NSF PHY11-25915 and by DOE grant DE-FG02-94ER40823. Research of PK at the Perimeter Institute is supported by the Government of Canada through Industry Canada and by the Province of Ontario through the Ministry of Economic Development and Innovation.

\bibliography{cpn1}
\bibliographystyle{nb}

\end{document}